\documentclass{article}

\usepackage{graphicx}
\usepackage{multirow}
\usepackage{amssymb} 
\usepackage{amsthm}
\usepackage{amsmath}
\usepackage{fullpage}
\usepackage[longnamesfirst]{natbib}
\usepackage{enumerate}
\usepackage{amsfonts, bm, amsmath, color, natbib, rotating, amsthm, subcaption, lscape, multicol, epstopdf, verbatim, hyperref}  
\usepackage{authblk}
\usepackage{xr}
\bibpunct{(}{)}{;}{a}{,}{,}

\usepackage{setspace}
\theoremstyle{plain}  \newtheorem{proposition}{Proposition}  

\theoremstyle{definition}    

\theoremstyle{remark}

\usepackage{xr}
\begin{document}

\newif\ifblinded

\title{Integer-Valued Functional Data Analysis for Measles Forecasting}

\ifblinded
\author{}
\else

\author{Daniel R. Kowal\thanks{Assistant Professor, Department of Statistics, Rice University, Houston, TX 77251-1892 (E-mail: \href{mailto:daniel.kowal@rice.edu}{daniel.kowal@rice.edu}).}}

\fi

\maketitle
\large 


\begin{abstract}
Measles presents a unique and imminent challenge for epidemiologists and public health officials: the disease is highly contagious, yet vaccination rates are declining precipitously in many localities. Consequently, the risk of a measles outbreak continues to rise. To improve preparedness, we study historical measles data both pre- and post-vaccine, and design new methodology to forecast measles counts with uncertainty quantification. We propose to model the disease counts as an integer-valued functional time series: measles counts are a \emph{function} of time-of-year and \emph{time-ordered} by year. The counts are modeled using a negative-binomial distribution conditional on a real-valued latent process, which accounts for the overdispersion observed in the data. The latent process is decomposed using an \emph{unknown} basis expansion, which is learned from the data, with dynamic basis coefficients. The resulting framework provides enhanced capability to model complex seasonality, which varies dynamically from year-to-year, and offers improved multi-month ahead point forecasts and substantially tighter forecast intervals (with correct coverage) compared to existing forecasting models. Importantly, the fully Bayesian approach provides well-calibrated and precise uncertainty quantification for epi-relevent features, such as the future value and time of the peak measles count in a given year. An \texttt{R} package is available online.  
\end{abstract}

\noindent {\bf KEYWORDS: Bayesian modeling; disease; prediction; MCMC; public health; time series}
\clearpage

\setcounter{page}{1}

\clearpage
\section{Introduction}
Forecasting the spread of infectious diseases is a fundamental goal of epidemiologists and healthcare suppliers. Accurate weeks- and months-ahead forecasts provide essential information for planning and allocation of medical resources and contribute to an informed and healthy community. Precise uncertainty quantification of future disease counts greatly aids medical and public health officials, and can save human lives. Other epi-relevant features, such as the peak number of cases and the time at which the peak occurs, may be equally important to forecast \citep{tabataba2017framework}. Indeed, despite the wide availability of vaccines for many infectious diseases in the United States, certain disease incidences are on the rise in many states, and often accompanied by fatalities \citep{van2013contagious}. 

\begin{figure}[h]
\begin{center}
\includegraphics[width=1\textwidth]{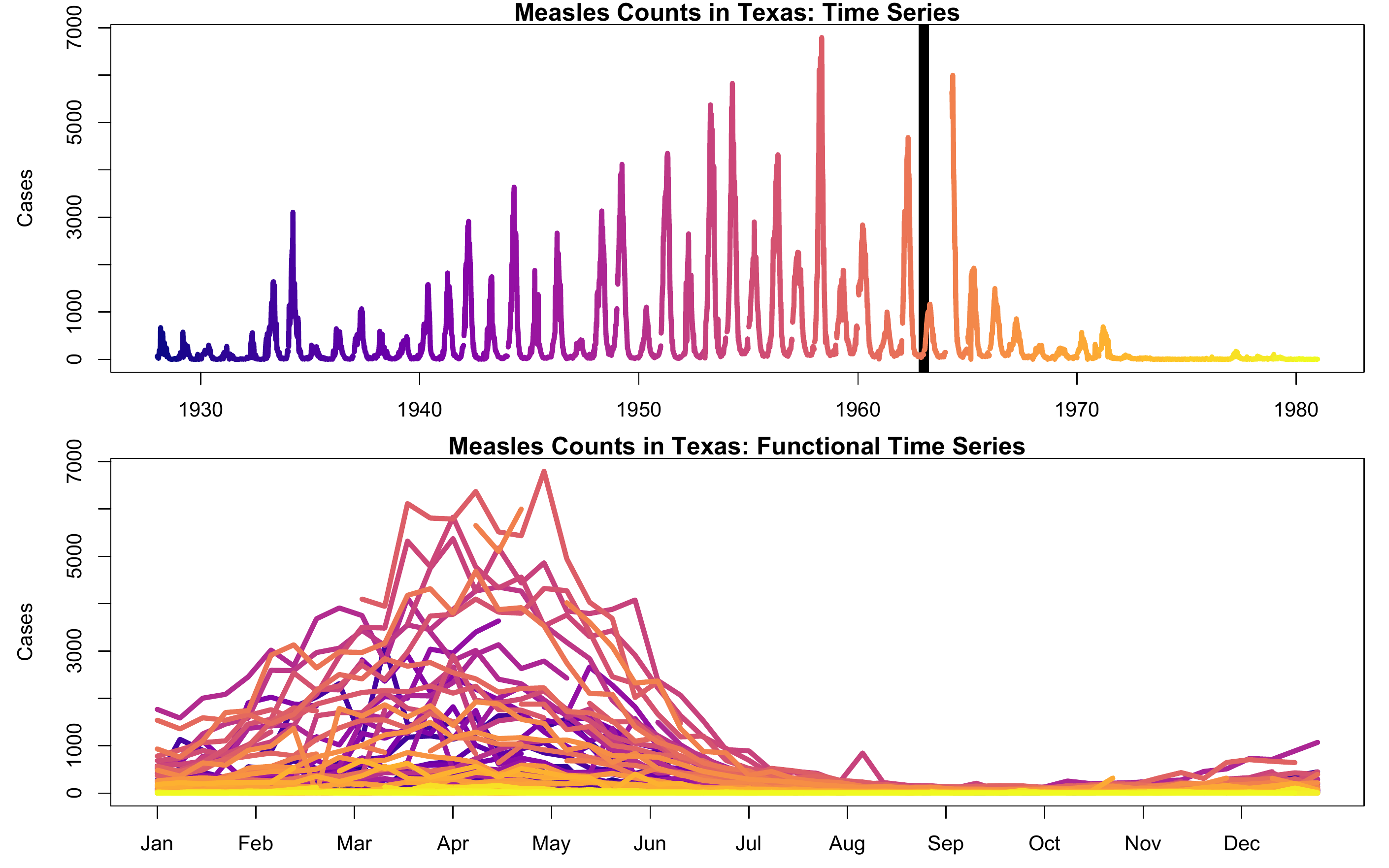}
\caption{Measles counts in Texas displayed as a time series ({\bf top}) and a functional time series ({\bf bottom}). As a functional time series, the weekly counts are viewed as a function of week-of-year with functions time-ordered by year. Each year corresponds to a single color which is aligned between the plots. The vertical bar (top) denotes the introduction of the first measles vaccine in 1963. The data are distinctly seasonal, with maximal variability between March and June. The plots only include data up to 1980 for display purposes. This figure appears in color in the electronic version of this article, and color refers to that version. 
 \label{fig:counts}}
\end{center}
\end{figure}

Among infectious diseases, measles presents an important and imminent challenge. Measles infections commonly occur among school-aged children, with incidence patterns that demonstrate complex seasonality (see Figure \ref{fig:counts}): in endemic localities, distinct intra-year seasonality persists both during and between epidemics, yet seasonality dissipates as the disease is eradicated \citep{durrheim2014measles}. Accounting for intra-year seasonality---which may evolve or vanish over time---is therefore essential for measles modeling and forecasting. Importantly, measles is the most infectious communicable disease known: an infected individual  in a susceptible population generates 12-18 new cases on average \citep{durrheim2014measles}. As a result, vaccination rates must be exceptionally high to prevent measles outbreaks.

The state of Texas is notably problematic: it is second most populous state in the United States, with a rapidly declining measles vaccination rate. Texas state law allows for ``reasons of conscience" (i.e., nonmedical) exemptions from immunization. The Texas Department of State Health Services reports that 1.07\% of students---or around 50,000 children---were granted exemptions in 2018, far above the 0.45\% who were granted exemptions in 2010 \citep{texasExempt}. \cite{lo2017public} carefully articulate the dangers of lapses in measles vaccination coverage in Texas, noting that consequences extend beyond school children and may adversely affect unvaccinated infants and adults, with implications for both public health and the statewide economy. The perilous combination of (i) a highly infectious disease and (ii) an expanding  population of susceptible, nonvaccinated children is particularly concerning, and leads experts to warn of a serious and imminent risk for a measles outbreak in Texas in the coming years \citep{hotez2016texas}.

To improve preparedness for a measles outbreak, we develop new methodology that provides accurate weeks- and months-ahead forecasts with precise uncertainty quantification. Within a Bayesian framework, we model measles counts as \emph{integer-valued functional data}: measles counts are a function of time-of-year with replicates (time-ordered) across years. This approach, illustrated in Figure \ref{fig:counts}, allows for flexible modeling of the intra-year seasonality, which varies dynamically from year-to-year---an important feature for modeling measles incidence. Here, the  intra-year seasonality pattern is modeled as unknown and learned from the data, which provides the  foundation for accurate long-term forecasting coupled with well-calibrated prediction intervals. The measles counts are modeled using a negative-binomial distribution conditional on a real-valued latent process, which offers several advantages: (i) it is well-defined for count data, 
 which is important for modeling small counts that occur during low seasons and especially post-vaccine;  (ii) it accounts for overdispersion observed in the data; (iii) it is compatible with an efficient Gibbs sampling algorithm for posterior inference; and (iv) by conditioning on a real-valued process, it provides a general framework for building upon the multitude of successful models for real-valued functional data. 

Prediction of disease incidence has become an important statistical research topic in recent years \citep{unkel2012statistical,nsoesie2014systematic}. Successful models commonly assume \emph{dynamic} and \emph{integer-valued} distributions for disease counts. The susceptible-infected-recovered (SIR) model, which specifies a mechanistic model for the spread of infectious disease via differential equations, corresponds to a negative-binomial model with latent dynamics \citep{bjornstad2002dynamics,dalziel2016persistent}. For influenza forecasting, \cite{osthus2018dynamic} improve upon the parametric limitations of the SIR model by  incorporating additional local dynamics via a reverse-random walk model. \cite{martinez2017bayesian} model dengue disease counts using a dynamic Poisson model, which includes covariates. These methods, including more classical seasonal time series models \citep{shumway2000time}, focus primarily on local week-to-week time-dependence, with limited flexibility for intra-year seasonalities. Alternatively, \cite{brooks2015flexible} propose an empirical Bayesian approach for modeling influenza counts as smooth functions of week-of-year. The aggregate success of these methods serves as motivation for the proposed Bayesian integer-valued functional data model.

The paper is organized as follows. 
The integer-valued functional data model is defined in Section \ref{methods}, with an accompanying MCMC algorithm for posterior inference in Section \ref{mcmc}. The  measles forecasting study is in Section \ref{measles}, which also includes in-sample inference for the vaccine effect. A simulation study is in Section \ref{sims}. We conclude in Section \ref{discussion}. 

\section{Modeling Integer-Valued Functional Data}\label{methods}
\subsection{The Model}
Suppose we observe (non-negative) integer-valued functions $Z_i\! : \mathcal{T} \to \mathcal{N} = \{0, 1,\ldots,\infty\}$ for $i=1,\ldots,n $ on some domain $\mathcal{T} \in \mathbb{R}^D$ for $D \in \mathbb Z^+$. For measles incidence data, $Z_i(\tau)$ corresponds to the measles counts in week $\tau$ of year $i$. The intra-year index $\tau \in \mathcal{T}$ is continuous (with $D=1$), so the model is well-defined for finer time scales, such as daily counts. The integer-valued functional data are modeled using a negative-binomial distribution conditional on a real-valued latent process, $\theta_i \! : \mathcal{T} \to \mathbb{R}$, and a scalar dispersion, $r > 0$:
\begin{equation}\label{nb-model}
[Z_i(\tau) | r, \theta_i(\cdot)] \stackrel{indep}{\sim} \mbox{NB}\left\{r,  \frac{\exp\left[\theta_i(\tau)\right]}{r + \exp\left[\theta_i(\tau)\right]}\right\}, \quad \tau \in \mathcal{T}
\end{equation}
The dispersion $r$ provides distributional flexibility, and in particular may incorporate overdispersion of the integer-valued data $Z_i(\cdot)$, while the real-valued process $\theta_i(\cdot)$ accounts for functional and other dependence. Explicitly, the negative-binomial density of $z \sim \mbox{NB}(r,\pi)$ for $r > 0$ and $\pi \in [0,1]$ is 
$ p(z | r, \pi) = \Gamma(r + z)/\left[\Gamma(r)z!\right]\pi^z (1-\pi)^r$ supported on $z \in \mathcal{N}
$. 
The conditional expectation and variance of $Z_i(\tau)$ in \eqref{nb-model} are related to $r$ and $\theta_i(\cdot)$ as follows:
 \begin{align}
\label{cond-mean}
\mathbb{E}[Z_i(\tau) | r, \theta_i(\cdot)] &= \exp\left[\theta_i(\tau)\right]\\
\label{cond-var}
\mbox{Var}[Z_i(\tau) | r, \theta_i(\cdot)] &= \exp\left[\theta_i(\tau)\right]\left\{1 + \exp\left[\theta_i(\tau)\right]/r\right\}
\end{align}
The process $\theta_i(\cdot)$ completely determines the conditional expectation of the integer-valued functional data $Z_i(\cdot)$, while the conditional variance exceeds the conditional expectation with greater overdispersion as $r >0$ decreases. Alternatively, model \eqref{nb-model} may be derived via the Poisson-Gamma hierarchical model $[Z_i(\tau) | \lambda_i(\tau) ]\stackrel{indep}{\sim}\mbox{Poisson}\left[\lambda_i(\tau)\right]$ with 
$[\lambda_i(\tau) | r, \theta_i(\cdot)] \sim \mbox{Gamma}\left\{r, r \exp\left[\theta_i(\tau)\right]\right\}$ and marginalizing over $\lambda_i(\tau)$. In the limit $r\rightarrow \infty$, \eqref{nb-model} converges to a Poisson distribution with mean parameter $ \exp\left[\theta_i(\tau)\right]$.

 $Z_i(\tau)$ is modeled as conditionally independent in \eqref{nb-model}, with all  functional and other dependence, such as time dependence, relegated to the real-valued process  $\theta_i(\tau)$. Importantly, the vast majority of Bayesian models for functional data are designed for real-valued processes, which suggests a wide range of options for $\theta_i(\cdot)$. We model the real-valued process using a functional basis expansion with Gaussian innovations:
\begin{align}
\label{process}
\theta_i(\tau)  &= \mu_i(\tau) + \epsilon_i(\tau), \quad [\epsilon_i(\cdot) | \sigma_{\epsilon}^2] \stackrel{indep}{\sim} N(0, \sigma_\epsilon^2) \\
\label{fdlm}
\mu_i(\tau) &= \sum_{k=1}^K f_k(\tau) \beta_{k,i}
\end{align}
where $\mu_i(\tau)$ is the conditional expectation of $\theta_i(\tau)$, $\epsilon_i(\cdot)$ is a white noise process, $f_k\!: \mathcal{T}\to \mathbb{R}$ for $k=1,\ldots,K$ are (known or unknown) basis functions, and $\{\beta_{k,i}\}$ are the corresponding basis coefficients. Model \eqref{process}-\eqref{fdlm} is designed to reproduce the classical setting for functional data analysis, where a real-valued function $\theta_i(\tau)$ is modeled as noisy observations of a smooth function $\mu_i(\tau)$, commonly via a basis expansion. For maximal flexibility, we model the basis functions $\{f_k\}$ as smooth yet unknown functions (subject to identifiability constraints), which produces a data-adaptive functional basis  \citep{kowal2018dynamic,kowal2018bayesian}. For modeling measles counts, learning $\{f_k\}$ corresponds to learning the intra-year seasonalities, with year-specific weights given by the coefficients $\{\beta_{k,i}\}$ (see Section \ref{flcs}).

The conditional expectation of $Z_i$ in \eqref{cond-mean} is non-smooth, since $\theta_i$ contains a white noise component $\epsilon_i$. A smoother representation is obtainable by marginalizing over $\epsilon_i$: 
\begin{equation}\label{cond-mean-2}
\mathbb{E}[Z_i(\tau) |  \mu_i(\cdot), \sigma_\epsilon, r] = \exp\left[\mu_i(\tau)\right]\exp(\sigma_e^2/2)
\end{equation}
where $\mu_i$ is smooth. Equation \eqref{process} may also include an additive offset or exposure, say $E_i(\tau) > 0$, on the log-scale such that $\mathbb{E}[Z_i(\tau) |  \mu_i(\cdot), \sigma_\epsilon] = \exp\left[\mu_i(\tau) + \log E_i(\tau)\right]\exp(\sigma_e^2/2) = E_i(\tau) \exp\left[\mu_i(\tau) \right]\exp(\sigma_e^2/2)$. In Section \ref{measles}, we include an offset for the population of Texas. 

The combination of a negative-binomial model \eqref{nb-model} and a Gaussian model \eqref{process} has appeared previously in other contexts. In place of the basis expansion \eqref{fdlm},  \cite{davis2009negative} propose a linear time series model, \cite{zhou2012lognormal} use a regression model, and \cite{klami2015polya} introduce a factor model, while each emphasizes distributional flexibility, in particular for overdispersion. The model framework of \eqref{nb-model},  \eqref{process}, and \eqref{fdlm} is compatible with a computationally efficient and convenient blocking structure for a Gibbs sampler, as in \cite{zhou2012lognormal} and \cite{klami2015polya}: $\theta_i(\cdot)$ is sampled from a full conditional Gaussian distribution using a P{\'o}lya-Gamma data augmentation scheme \citep{polson2013bayesian}, and conditional on $\theta_i(\cdot)$, model \eqref{process}-\eqref{fdlm} is simply a Gaussian functional data model with ``data" $\theta_i(\cdot)$ and therefore may utilize existing samplers for $\sigma_\epsilon$ and the parameters comprising $\mu_i(\cdot)$.

Model \eqref{fdlm} may be accompanied by various functional data models, such as function-on-scalars regression \citep{goldsmith2015generalized,goldsmith2016assessing}, functional multi-level and mixed models \citep{morris2006wavelet,di2009multilevel,zhu2011robust}, functional autoregressive models for time-ordered functional data \citep{kowal2017functional}, among others (e.g., \citealp{silverman2005functional,morris2015functional}). To incorporate time-dependence from year-to-year, we consider the following autoregressive model:
\begin{equation}\label{far}
\beta_{k,i} = \mu_k + \phi_k(\beta_{k,i-1} - \mu_k) + \eta_{k,i}, \quad  \eta_{k,i} \stackrel{indep}{\sim}N(0, \sigma_{\eta_{k,i}}^2)
\end{equation}
Generalizations for multiple lags and vector autoregressions are available, yet less parsimonious. The model properties of \eqref{process}, \eqref{fdlm}, and \eqref{far} are discussed in Section \ref{properties}.

The priors for $\mu_k$ and $\eta_{k,i}$ control the shrinkage behavior of the model for $\beta_{k,i}$. Without adequate shrinkage on $\mu_k$ and $\eta_{k,i}$, the model may be overly sensitive to the number of basis functions $K$. To mitigate the impact of the choice of $K$, we introduce ordered shrinkage across coefficients $k=1,\ldots,K$ via a multiplicative gamma process (MGP) prior \citep{bhattacharya2011sparse} for $\mu_k$ and $\eta_{k,i}$. The resulting variance for each of $\mu_k$ and $\eta_{k,i}$ is stochastically decreasing in $k$, so coefficients become \emph{a priori} less important for larger $k$. Compared to alternative approaches that treat $K$ as unknown \citep{suarez2017bayesian}, the MGP approach is computationally scalable and does not require complex and intensive computing procedures such as reversible jump MCMC. Specifically, the priors are  $\mu_k \stackrel{indep}{\sim}N(0, \sigma_{\mu_k}^2)$ with $\sigma_{\mu_k}^{-2} = \prod_{\ell \le k} \delta_{\mu_\ell}$, $ \delta_{\mu_1} \sim \mbox{Gamma}(a_{\mu_1}, 1)$, $ \delta_{\mu_\ell} \sim \mbox{Gamma}(a_{\mu_2}, 1)$ for $\ell > 1$, and $\sigma_{\eta_{k,i}}^2 = \sigma_{\eta_k}^2/\xi_{\eta_{k,i}}$ with $\sigma_{\eta_k}^{-2} = \prod_{\ell \le k} \delta_{\eta_\ell}$, $ \delta_{\eta_1} \sim \mbox{Gamma}(a_{\eta_1}, 1)$, $ \delta_{\eta_\ell} \sim \mbox{Gamma}(a_{\eta_2}, 1)$ for $\ell > 1$, and $\xi_{\eta_{k,i}} \stackrel{iid}{\sim} \mbox{Gamma}(\nu_\eta/2, \nu_\eta/2)$ with $\nu_\eta \sim \mbox{Uniform}(2, 128)$ to induce heavier tails in the marginal distribution of $\eta_{k,i}$. The hyperpriors  $a_{\mu_1}, a_{\mu_2}, a_{\eta_1},a_{\eta_2} \stackrel{iid}{\sim}\mbox{Gamma}(2,1)$ allow the data to determine the rate of ordered shrinkage. For the autoregressive coefficient in \eqref{far}, we assume the prior $\left(\phi_k + 1\right)/2 \stackrel{iid}{\sim}\mbox{Beta}(a_\phi,b_\phi)$ to constrain $|\phi_k| < 1$ for stationarity, and select $a_\phi = 5$ and $b_\phi = 2$ for measles and simulated data. These choices were successfully applied for Gaussian functional data in \cite{kowal2018dynamic} and \cite{kowal2018bayesian}.

\subsection{Modeling the Basis Functions}\label{flcs}
While many options exist for the basis functions $f_k(\cdot)$ in \eqref{fdlm}, we follow the highly flexible approach of \cite{kowal2017bayesian} for \emph{unknown} $f_k$ with the computational improvements in \cite{kowal2018dynamic} and \cite{kowal2018bayesian}. Modeling each basis function $f_k$ as unknown (i) produces a data-adaptive functional basis and (ii) incorporates the uncertainty about $f_k$ into the posterior distribution. Specifically, we let $f_k(\tau) = \bm b'(\tau) \bm \psi_k$ for $\bm b'(\tau) = (b_1(\tau), \ldots, b_{L_m}(\tau))$ a vector of low-rank think plate splines and  $\bm\psi_k$ a vector of unknown coefficients. Low-rank thin plate splines are flexible, smooth, and typically efficient within MCMC samplers \citep{crainiceanu2005bayesian}, and are well-defined for $\mathcal{T}\subset \mathbb{R}^D$ with $D \in \mathbb{Z}^+$. For smoothness, we assume the prior  $\bm \psi_k \sim N(\bm 0, \lambda_{f_k}^{-1} \bm \Omega^{-1})$, where $\bm \Omega$ is a $L_m \times L_m$ known roughness penalty matrix, such as $\bm \Omega = \int \ddot{\bm b}(\tau)\ddot{\bm b}'(\tau) d\tau$ for $\ddot{\bm b}$ the second derivative of $\bm b$, and $\lambda_{f_k} > 0$ is an unknown smoothing parameter. Details on the construction of $\bm b(\cdot)$ and $\bm \Omega$, including useful reparametrizations, are given in \cite{kowal2018dynamic}. For identifiability, we enforce the matrix orthonormality constraint $\bm F' \bm F = \bm I_K$, where $\bm F = (\bm f_1, \ldots, \bm f_K)$ is the $m \times K$ basis matrix and $\bm f_k = (f_k(\tau_1),\ldots, f_k(\tau_m))'$ is the basis function $f_k$ evaluated at the observation points $\tau_1,\ldots,\tau_m$. The matrix orthonormality constraint, coupled with the ordering implied by the MGP prior, is sufficient for identifiability, and may be leveraged to improve computationally efficiency in sampling the coefficients $\{\beta_{k,i}\}$ as in \cite{kowal2018dynamic} and \cite{kowal2018bayesian}.

\subsection{Properties of the Model}\label{properties}
To understand the implications of a functional time series model for seasonal time series data, we consider the covariance properties of the real-valued process $\theta_i(\tau)$ implied by \eqref{process}-\eqref{fdlm}. Naturally, the properties of $\theta_i(\tau)$ are important for the integer-valued functional data $Z_i(\tau)$, and for real-valued functional data we may simply replace $\theta_i(\tau)$ with the functional observations. Let  $\{\vartheta_t\}_{t=1}^T$ denote a seasonal time series with seasonality $m$ and $n$ seasons, so $T = nm$. To model the seasonal time series as a functional time series, we map $\theta_i(\tau) = \vartheta_{\tau + (i-1)m}$ for each $\tau = 1,\ldots, m$. In our application, the seasonality is $m=52$ weeks and the seasons are years $i=1,\ldots,n$. Consequently, the covariance of the process $\theta_i(\tau)$ directly determines the seasonal covariance of $\vartheta_t$. 

Let $\bm f'( \tau) = (f_1( \tau), \ldots, f_K(\tau))$  and $\bm \beta_i' = (\beta_{1,i}, \ldots, \beta_{K,i})$.

\begin{proposition}\label{contemp-covar}
The contemporaneous covariance function of $\theta_i(\tau)$ in \eqref{process}, conditional on  $\{f_k\}$, is full rank with
\begin{equation}\label{contempcovar}
\mbox{Cov}\left[\theta_i(\tau), \theta_{i}( u)\right]  =  \mbox{Cov}\left[\vartheta_{\tau + (i-1)m}, \vartheta_{u + (i - 1)m}\right] 
=\bm f' (\tau) \mbox{Cov}\left(\bm\beta_{i}\right) \bm f( u) + \mathbb{I}\{ \tau =  u\} \sigma_{\epsilon}^2
\end{equation} 
\end{proposition}
The covariance function in \eqref{contempcovar} is the covariance between weeks $\tau$ and $u$ of the same year $i$. If the coefficients $\{\bm \beta_i\}$ are stationary, then the covariance of $\bm \beta_i$ is constant and the within-year covariance of $\vartheta_t$ does not vary from year-to-year. The week-to-week dependence is primarily governed by the basis functions $\{f_k\}$, which are smooth, implying that the seasonal covariance in \eqref{contempcovar} is also smooth. For weeks in different years, we have the following:
\begin{proposition}\label{auto-covar}
The lag-$\ell$ autocovariance function of $\theta_i(\tau)$ in \eqref{process}, conditional on  $\{f_k\}$, is rank $K$ with
\begin{equation}\label{autocovar}
\mbox{Cov}\left[\theta_i(\tau), \theta_{i-\ell}( u)\right] 
=  \mbox{Cov}\left[\vartheta_{\tau + (i-1)m}, \vartheta_{u + (i-\ell - 1)m}\right] 
=\bm f' (\tau) \mbox{Cov}\left(\bm\beta_{i}, \bm \beta_{i-\ell}\right) \bm f( u)
\end{equation} 
\end{proposition}
For a seasonal time series,  \eqref{autocovar}  determines the covariance of $\vartheta_t$ between different seasons of different years. By modeling the dynamics of $\bm \beta_i$ (with respect to $i$), we may incorporate a flexible model for time-varying seasonality. In the special case of \eqref{far}, the covariance in \eqref{autocovar} simplifies to
\begin{equation}\label{autocovar-ar}
\mbox{Cov}\left[\theta_i(\tau), \theta_{i-\ell}( u)\right] =
 \sum_{k=1}^K \left[ 
f_k(\tau) f_k( u)\phi_k^\ell \sigma_{\eta_k}^2/\left(1-\phi_k^2\right)\right]
\end{equation} 
Seasonality is determined by the basis functions $f_k$, which are learned from the data, while year-to-year dependence is controlled by the autoregressive coefficients $\{\phi_k\}$. For the same week $\tau$ in different years $i$ and $i-\ell$,  \eqref{autocovar-ar} simplifies to $\mbox{Cov}\left[\theta_i(\tau), \theta_{i-\ell}( \tau)\right] =
 \sum_{k=1}^K \left[ 
f_k^2(\tau)\phi_k^\ell \sigma_{\eta_k}^2/\left(1-\phi_k^2\right)\right]$. Naturally, the year-over-year covariance depends on the time-of-year via $f_k^2(\tau)$: for example, the seasonal behavior during the peak months in the late spring may differ from the seasonal behavior during low seasons in the fall  (see Figure \ref{fig:counts}).

\section{MCMC Sampling Algorithm}\label{mcmc}
We develop a computationally efficient MCMC sampling algorithm for model \eqref{nb-model}, \eqref{process}, and \eqref{fdlm}. A fundamental observation is that, conditional on $\theta_i(\tau)$, the remaining parameters in $\mu_i(\tau)$ and $\sigma_\epsilon$ may be sampled using MCMC methods for Gaussian functional data, such as \cite{morris2006wavelet}, \cite{di2009multilevel}, \cite{zhu2011robust}, \cite{kowal2018dynamic}, and \cite{kowal2018bayesian}. Therefore, the primary consideration is obtaining a sampler for the negative-binomial parameters $r$ and $\theta_i(\tau)$ in \eqref{nb-model}. For the dispersion parameter $r > 0$, we assume the half-Cauchy prior distribution $r \sim C^+(0, 10)$ and sample from the full conditional distribution using the univariate slice sampler \citep{neal2003slice}. 

To draw from the full conditional distribution of $\{\theta_i(\cdot)\}$, we employ a {{P}}{\'o}lya--{{G}}amma data augmentation scheme \citep{polson2013bayesian}, which produces a computationally efficient sampling algorithm that does not require any tuning.  The likelihood for $\theta_i(\tau)$ is proportional to $
p(Z_i(\tau) | r, \theta_i(\cdot)) \propto \left\{\exp\left[\theta_i(\tau) - \log r\right]\right\}^{Z_i(\tau)}/\left\{1 + \exp\left[\theta_i(\tau) - \log r\right]\right\}^{r + Z_i(\tau)}
$, which is proportional to a $Z$-distribution for $\left[\theta_i(\tau) - r\right]$. 
The likelihood for $\theta_i(\tau)$ is therefore proportional to the likelihood implied by $Z_i^{PG}(\tau) = \theta_i(\tau) + \nu_i(\tau)$ with $ Z_i^{PG}(\tau)  \equiv  \left[Z_i(\tau) - r\right]/\left[2\xi_{i}(\tau)\right] + \log r$ and $[\nu_i(\tau) | \xi_{i} ] \stackrel{indep}{\sim} N(0, \xi_{i}^{-1}(\tau))$ for $\xi_i(\tau) \stackrel{iid}{\sim}PG(Z_i(\tau) + r, 0)$ a P{\'o}lya-Gamma random variable \citep{kowal2017dynamic}. Sampling proceeds by drawing $\theta_i(\cdot)$ from its Gaussian full conditional distribution and the auxiliary  P{\'o}lya-Gamma random variables from the full conditional distribution $\xi_i(\tau) \stackrel{iid}{\sim} PG\left[Z_i(\tau) + r, \theta_i(\tau) - \log r\right]$. Additional background on P{\'o}lya-Gamma augmentation is available in \cite{zhou2012lognormal}, \cite{polson2013bayesian}, and \cite{klami2015polya}.

Let $\tau_1,\ldots,\tau_m$ denote the observation points of $Z_i(\cdot)$. While these points are assumed to be common for all $i=1,\ldots,n$, sparsely- or irregularly-sampled functional data may be accommodated via an imputation step. An outline of the Gibbs sampling algorithm is below:
\begin{enumerate}

\item {\bf Imputation:} Sample  $\left[Z_i(\tau^*) | \cdots\right] \stackrel{indep}{\sim} \mbox{NB}\left(r,  \exp\left[\theta_i(\tau^*)\right]/\left\{r + \exp\left[\theta_i(\tau^*)\right]\right\}\right)$  for all unobserved $Z_i(\tau^*)$; 

\item {\bf Dispersion:} Sample from $[r | \{Z_i(\cdot), \theta_i(\cdot)\}]$ using a slice sampler  \citep{neal2003slice}. 

\item {\bf Parameter Expansion:}  Sample $[\xi_i(\tau_j) | \cdots] \stackrel{indep}{\sim}PG\left[Z_i(\tau_j) + r, \theta_i(\tau_j) - \log r\right]$ using \cite{polson2013bayesian};

\item {\bf Process:} Sample $[\theta_i(\tau_j) | \cdots] \stackrel{indep}{\sim} N\left(Q_{\theta_{j,i}}^{-1} \ell_{\theta_{j,i}}, Q_{\theta_{j,i}}^{-1}\right)$ where $Q_{\theta_{j,i}} =\xi_{i}(\tau_j) + \sigma_\epsilon^{-2}$ and  $\ell_{\theta_{j,i}} = \xi_{i}(\tau_j) Z_i^{PG}(\tau_j) + \sigma_\epsilon^{-2} \mu_i(\tau_j)$  with $ Z_i^{PG}(\tau_j)  \equiv \left[Z_i(\tau) - r\right]/\left[2\xi_{i}(\tau)\right] + \log r$;

\item {\bf Rest:} given $\{\theta_i(\tau_j)\}$, sample the parameters of the Gaussian functional data model implied by \eqref{fdlm}, including $\sigma_\epsilon$, $\{\beta_{k,i}\}$, and $\{f_k\}$. 
\end{enumerate}
For model \eqref{far}, the final Gibbs block above uses the efficient sampler in \cite{kowal2018dynamic}. While we omit details for brevity, the \cite{kowal2018dynamic} sampler iteratively draws (i) the basis functions $\{f_k\}$ using a Bayesian backfitting sampler for Bayesian splines \citep{hastie2000bayesian}, (ii) the (dynamic) basis coefficients $\{\beta_{k,i}\}$ using a state space sampler \citep{durbin2002simple}, and (iii) the variance components $\sigma_\epsilon$ and the MGP shrinkage parameters with known full conditional distributions. An appealing feature of the proposed Gibbs sampler is its modularity: substituting other models in \eqref{fdlm} only impacts the final Gibbs block above, which may directly incorporate existing samplers for Gaussian functional data models. 

Draws from the posterior predictive distribution, say $\tilde Z_i(\tau^*)$, are generated using step 1. Inference on epi-relevant features proceeds by evaluating $h(\tilde Z_i)$ for each draw from the posterior predictive distribution, where $h$ computes the epi-relevant feature, such as the maximum count over $\tau^*$ in year $i$.

\section{Measles Counts in Texas}\label{measles}
We model and forecast weekly measles counts in Texas using publicly-available data from Project Tycho \citep{van2013contagious}. 
Data are available from 1928-2002 with approximately 10\% of counts missing, and exhibit substantial  overdispersion: the sample mean and sample variance of the observed counts are 355 and 628,746, respectively. Figure \ref{fig:counts} plots the observed counts, and demonstrates how we construct a functional time series using seasonal time series: the counts are re-organized as functions of week-of-year and time-ordered by year.  The annual seasonality is evident, with maximal variability between March and June. The amplitude varies from year-to-year, and vanishes altogether shortly following the introduction of the vaccine in 1963 \citep{plotkin2014history}. In more recent years, measles counts are comparatively low: from 2006-2016 there were 53 total cases, 27 of which occurred in 2013 alone \citep{texasMeasles}. However, as noted previously, declining vaccination rates in Texas imply a rapidly growing susceptible population, which continuously increases the risk of an outbreak.

\subsection{Measles Counts: Forecasting}\label{measles:fore}
For optimal preparedness and resource allocation, multi-weeks--ahead forecasts with uncertainty quantification are essential. However, adequate forecasting performance on post-vaccine data alone may be insufficient: in the event of a measles outbreak, counts may reflect pre-vaccine seasonality patterns and volume. Therefore, it is imperative that forecasting technologies be capable of performing adequately both pre- and post-vaccine.

Let $Z_i(\tau)$ denote the measles incident count for week $\tau = 1,\ldots, m = 52$ within year $i = 1,\ldots, n$. Equivalently, we may write the observed measles counts more traditionally as a time series of counts $z_t, t= 1,\ldots, T = nm$ with $Z_i(\tau) = z_{\tau + (i-1)m}$. To accompany the measles counts, we include annual population data for Texas \citep{texasPop}  as an offset $E_i(\tau) = E_i$ in \eqref{process}-\eqref{fdlm}, so \eqref{process} becomes $\theta_i(\tau) = \log E_i(\tau)  + \mu_i(\tau) + \epsilon_i(\tau)$, which implies the conditional expectation \eqref{cond-mean-2} may be expressed as a rate, $\mathbb{E}[Z_i(\tau)/E_i(\tau) |  \mu_i(\cdot), \sigma_\epsilon, r]  =  \exp\left[\mu_i(\tau)\right]\exp(\sigma_e^2/2)$. 

\subsubsection{Forecasting Design}
\begin{figure}[h]
\begin{center}
\includegraphics[width=.49\textwidth]{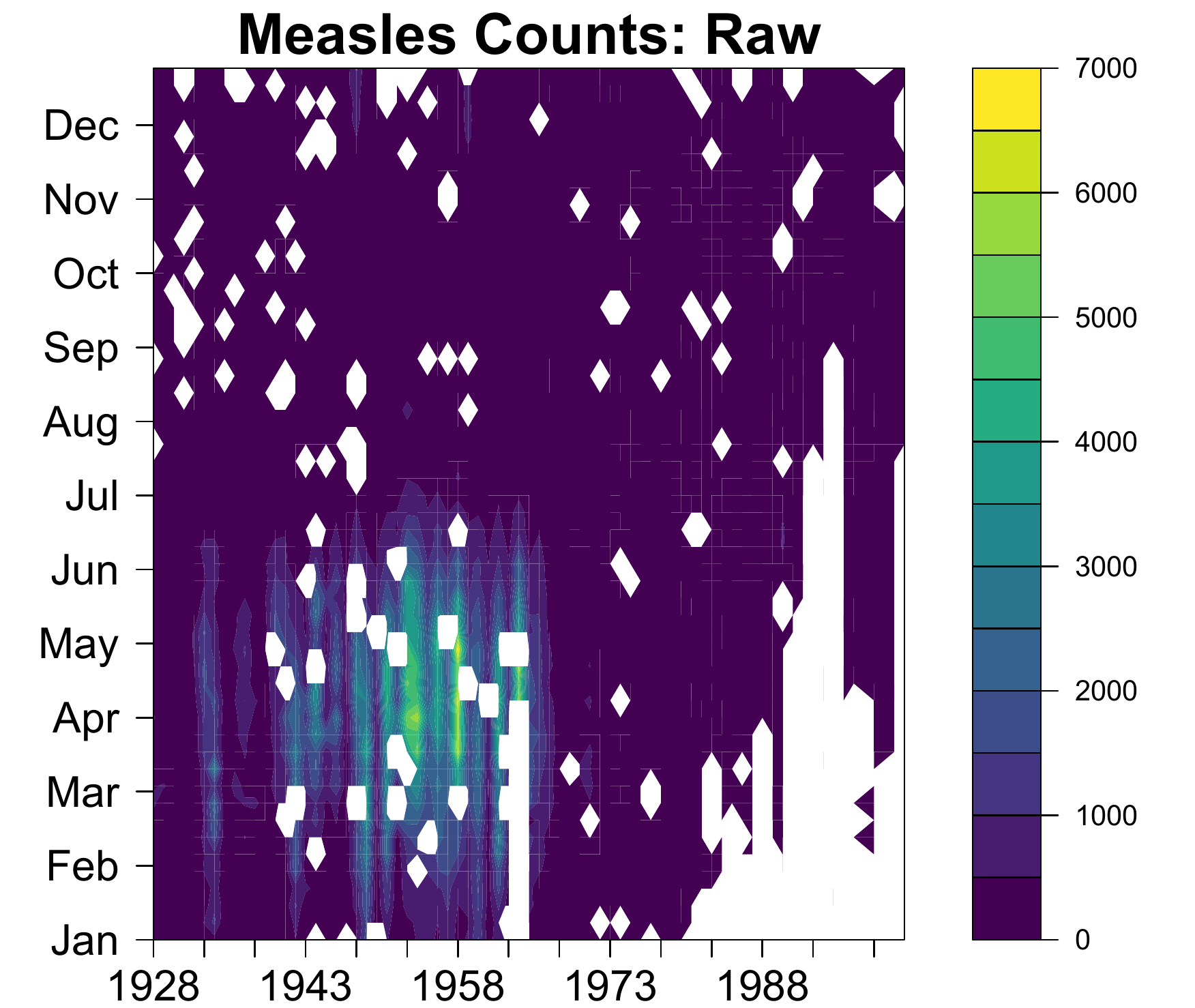}
\includegraphics[width=.49\textwidth]{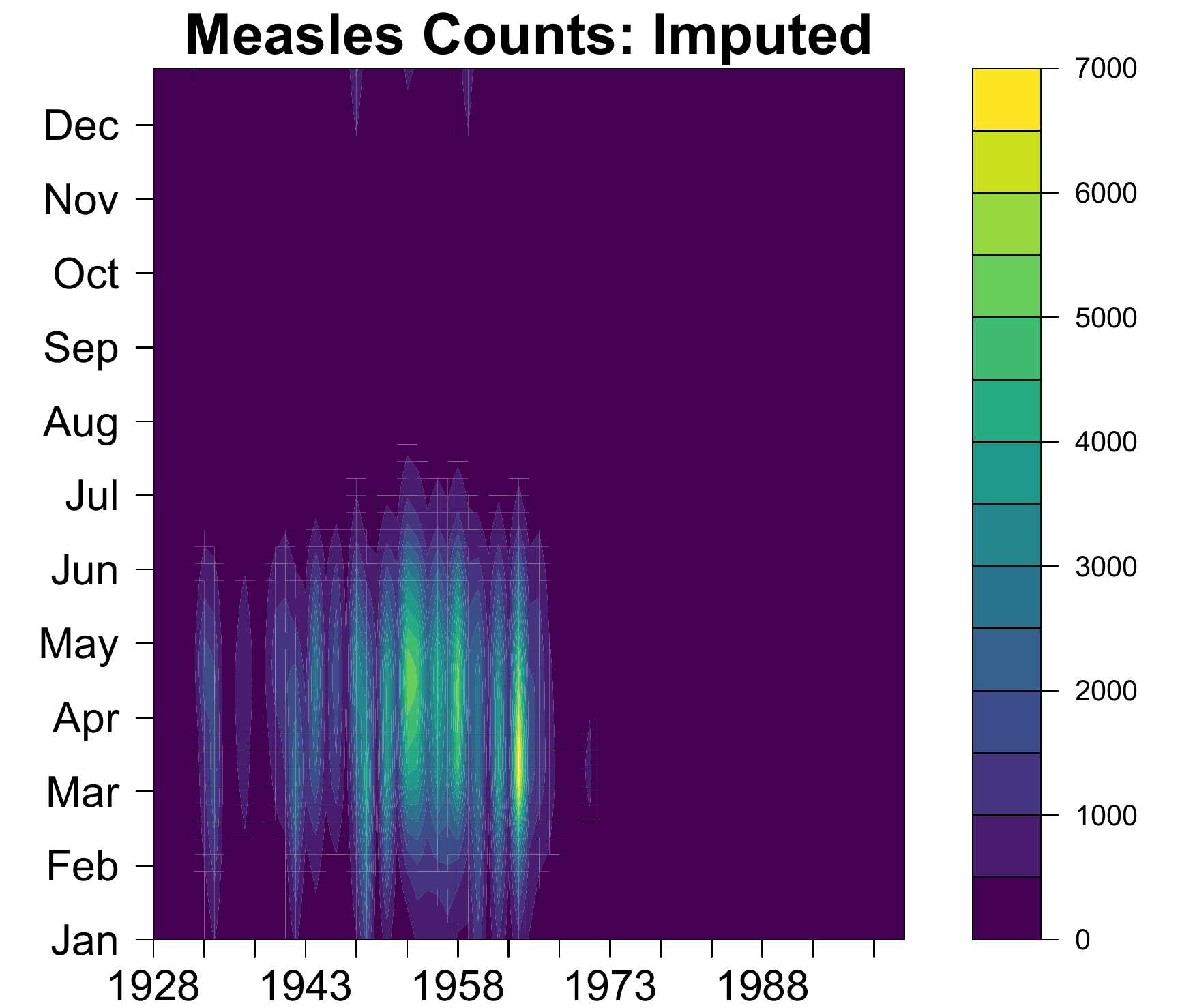}
\caption{Measles counts in Texas ({\bf left}) and the expected counts under the proposed model ({\bf right}). The 10\% missing data (white regions) are automatically imputed within the model. 
 \label{fig:impute}}
\end{center}
\end{figure}

For each year from 1950-1980, we compute multi-weeks-ahead forecasts  given (i) historical measles counts and population data from 1928 up to the forecasting year and (ii) measles counts and population data from the first $m_0$ weeks of the forecasting year. By observing a small number $m_0$ of weeks at the beginning of the year, the forecasting models may adapt to each year's distinct seasonality pattern in order to forecast the remaining $m-m_0$ weeks of the year. 
We consider $m_0 = 9$ weeks and $m_0 = 25$ weeks: the former case $m_0=9$ is more challenging, since it  constrains the partial observations almost entirely to January-February and therefore requires forecasts during the peak months of March-June as well as the subsequent decline. The forecasting period 1950-1980 includes pre- and post-vaccine years, and terminates in 1980 due to substantial missingness thereafter (see Web Figure \ref{fig:impute}). 

\begin{figure}[h]
\begin{center}
\includegraphics[width=.8\textwidth]{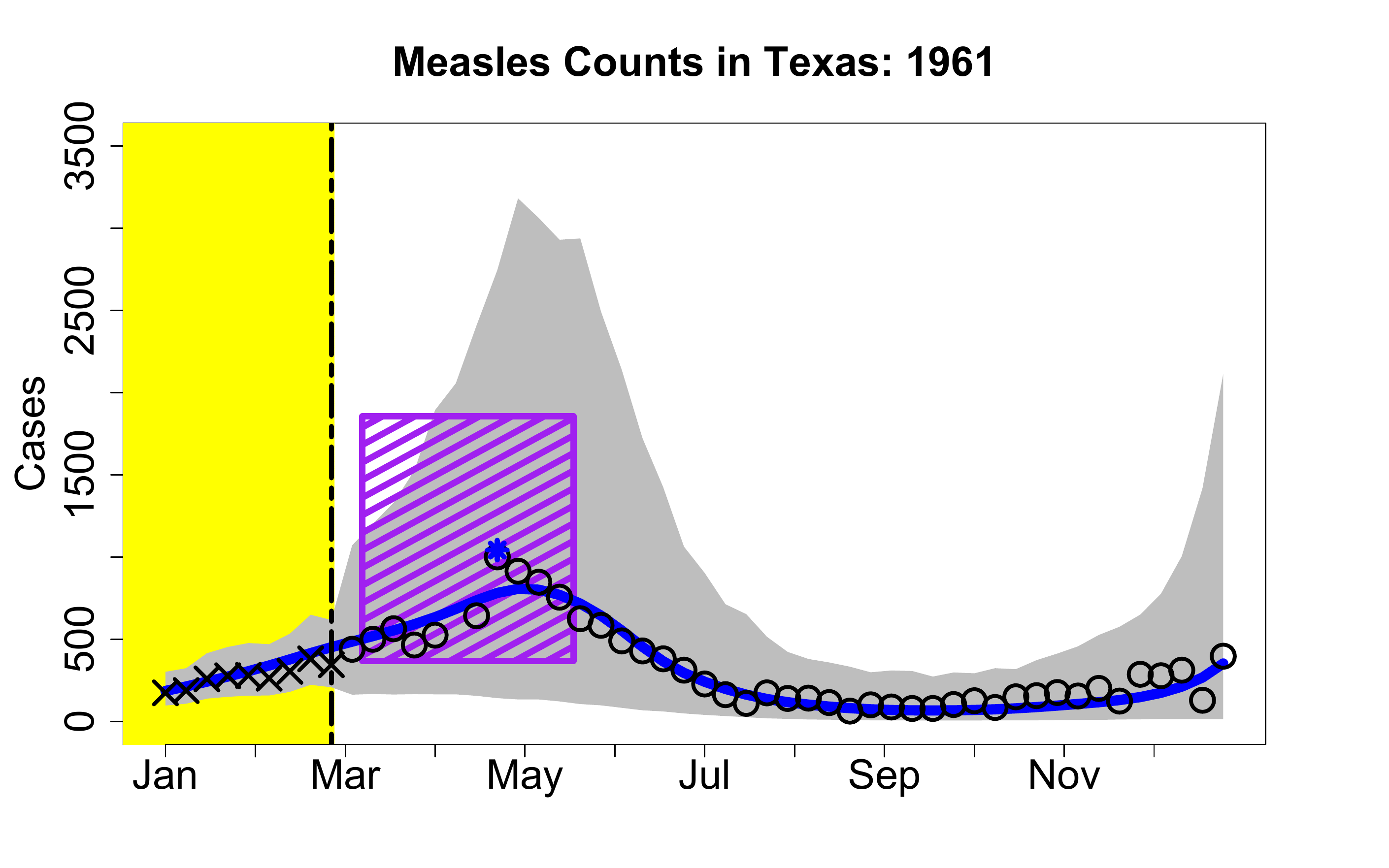}
\includegraphics[width=.4\textwidth]{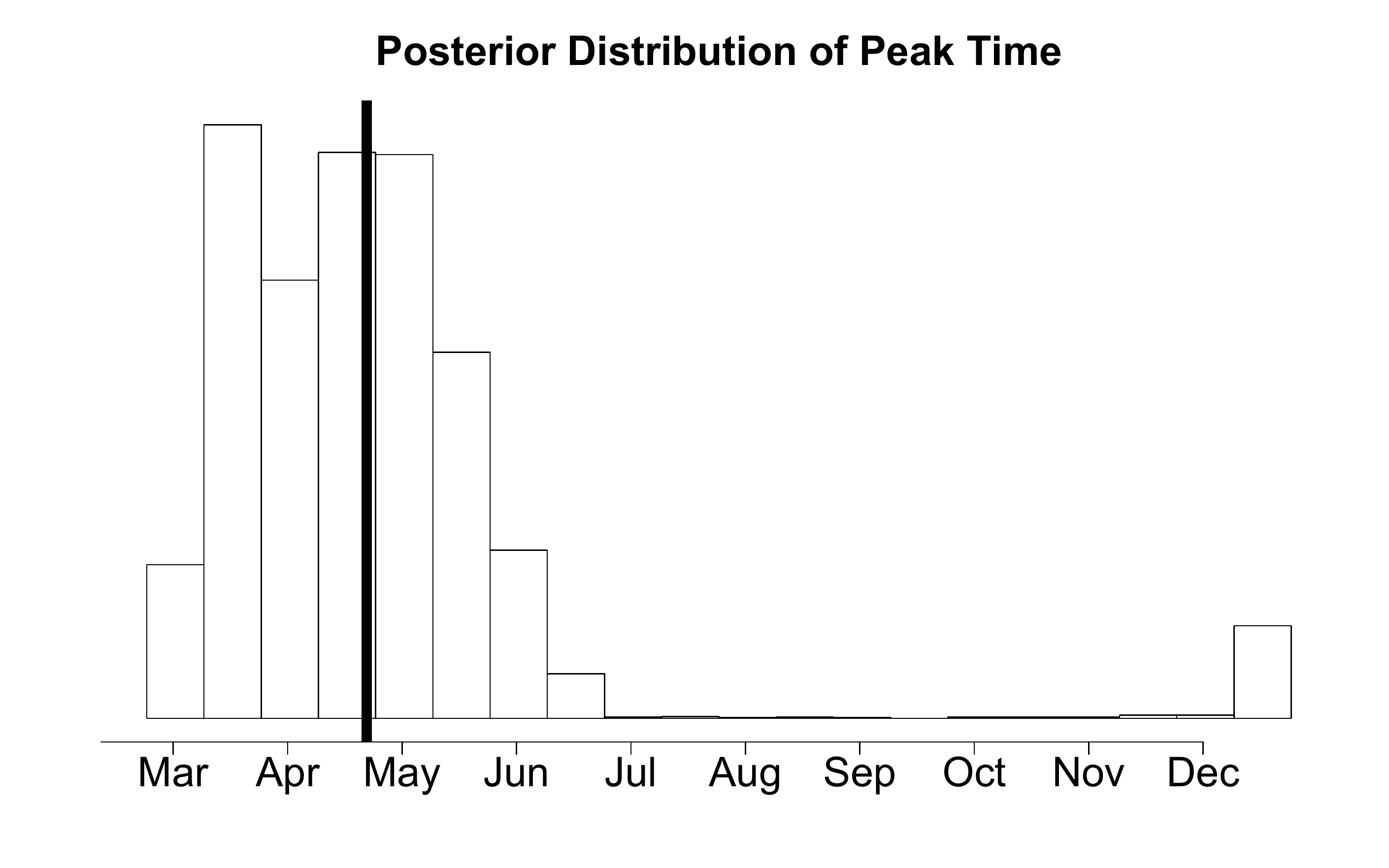}
\includegraphics[width=.4\textwidth]{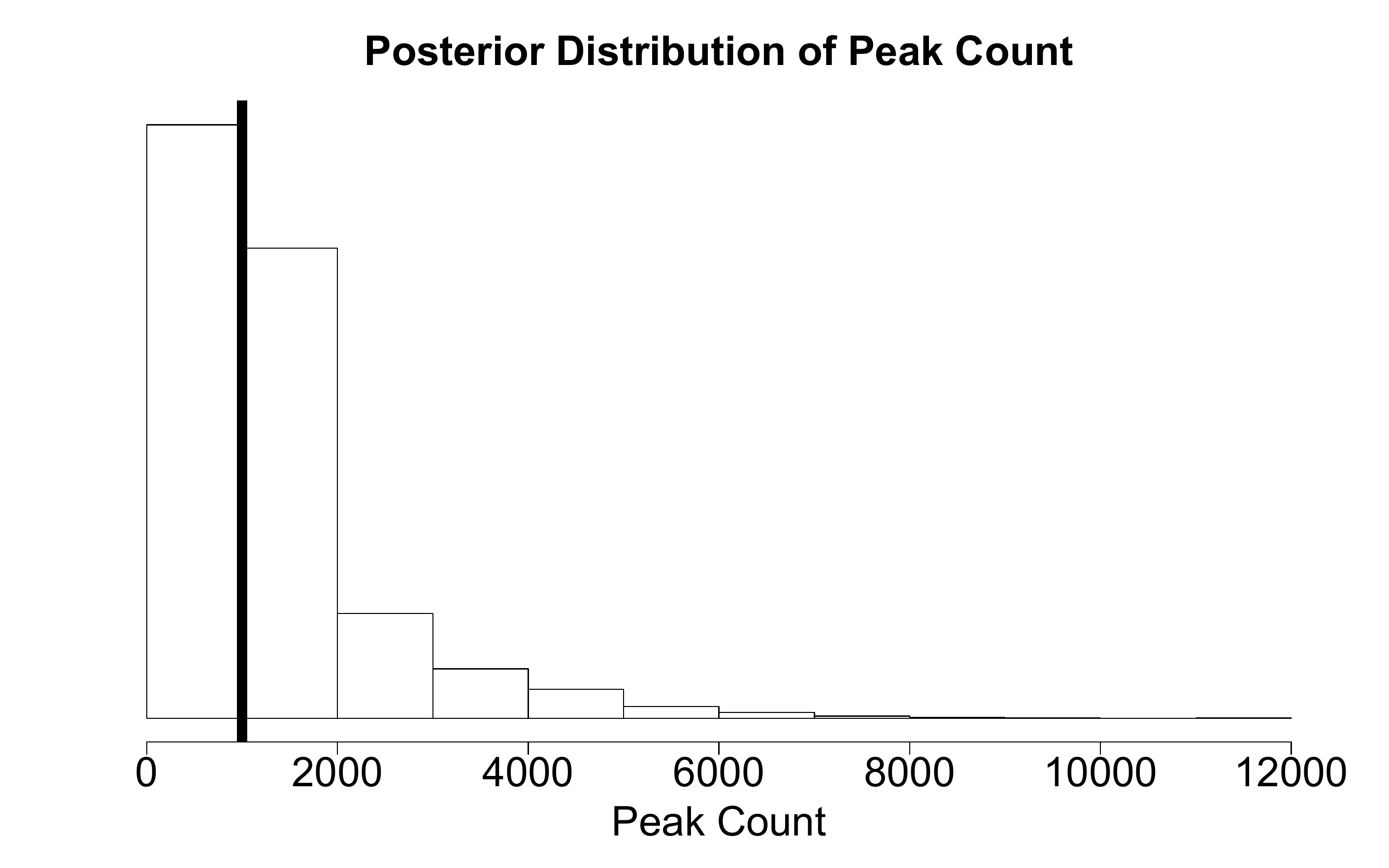}
\caption{Forecasting summary in 1961. {\bf Top:} Using data from the first 9 weeks (x-marks, yellow region), we forecast measles counts for the remainder of the year (circles). The posterior predictive mean (blue line) and 95\% intervals (gray region) provide a forecast with uncertainty quantification. The posterior predictive distribution provides inference for the peak time and peak counts: here, the posterior median time and peak count (blue star) and 80\% posterior credible intervals for the peak time and peak count (purple region). The posterior distribution for the peak time ({\bf bottom left}) and the posterior distribution for the peak count ({\bf bottom right}) are plotted together with the observed values (vertical black lines, {\bf bottom}). This figure appears in color in the electronic version of this article, and color refers to that version. 
 \label{fig:emp10-52}}
\end{center}
\end{figure}

\begin{figure}[h]
\begin{center}
\includegraphics[width=.8\textwidth]{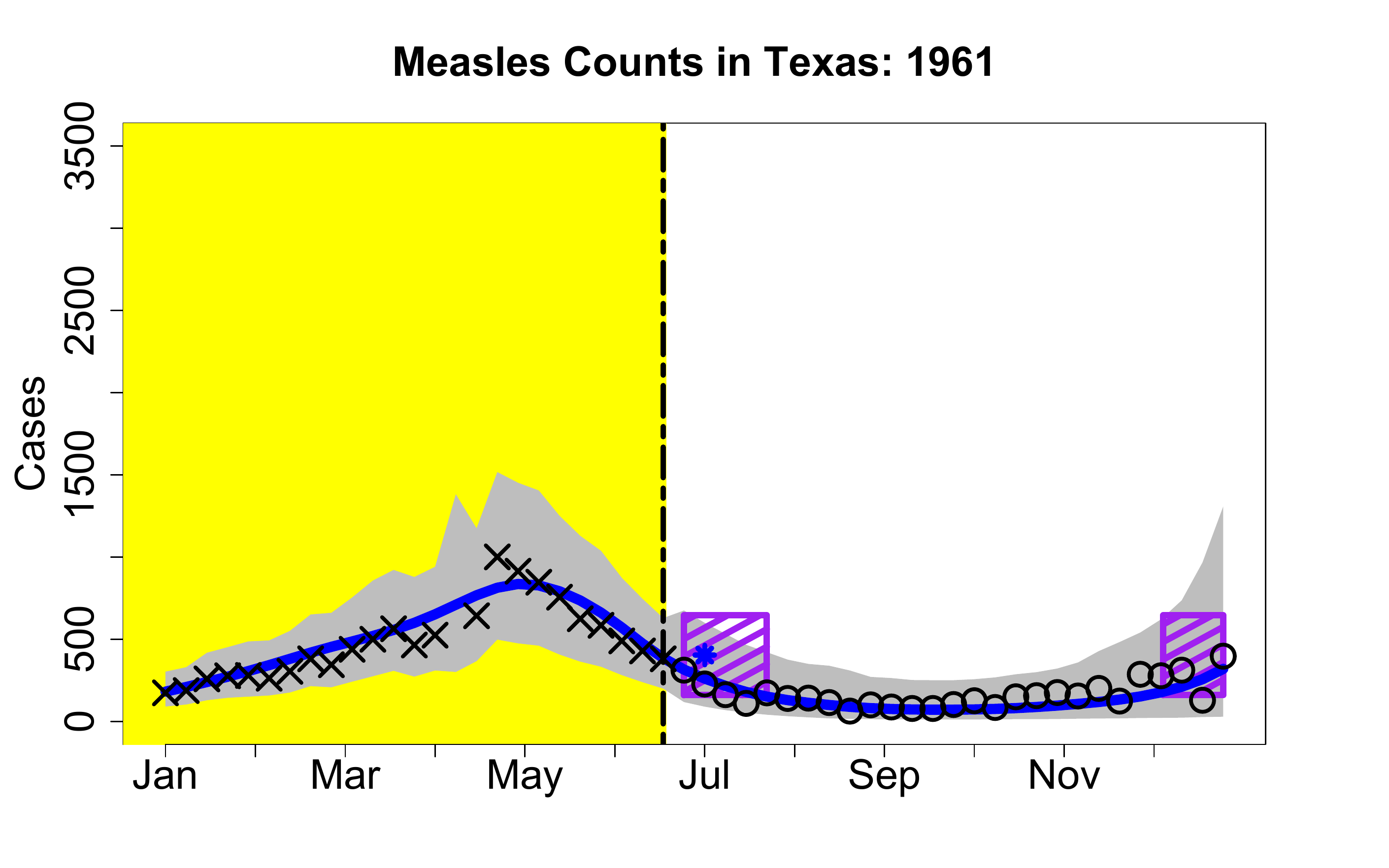}
\includegraphics[width=.4\textwidth]{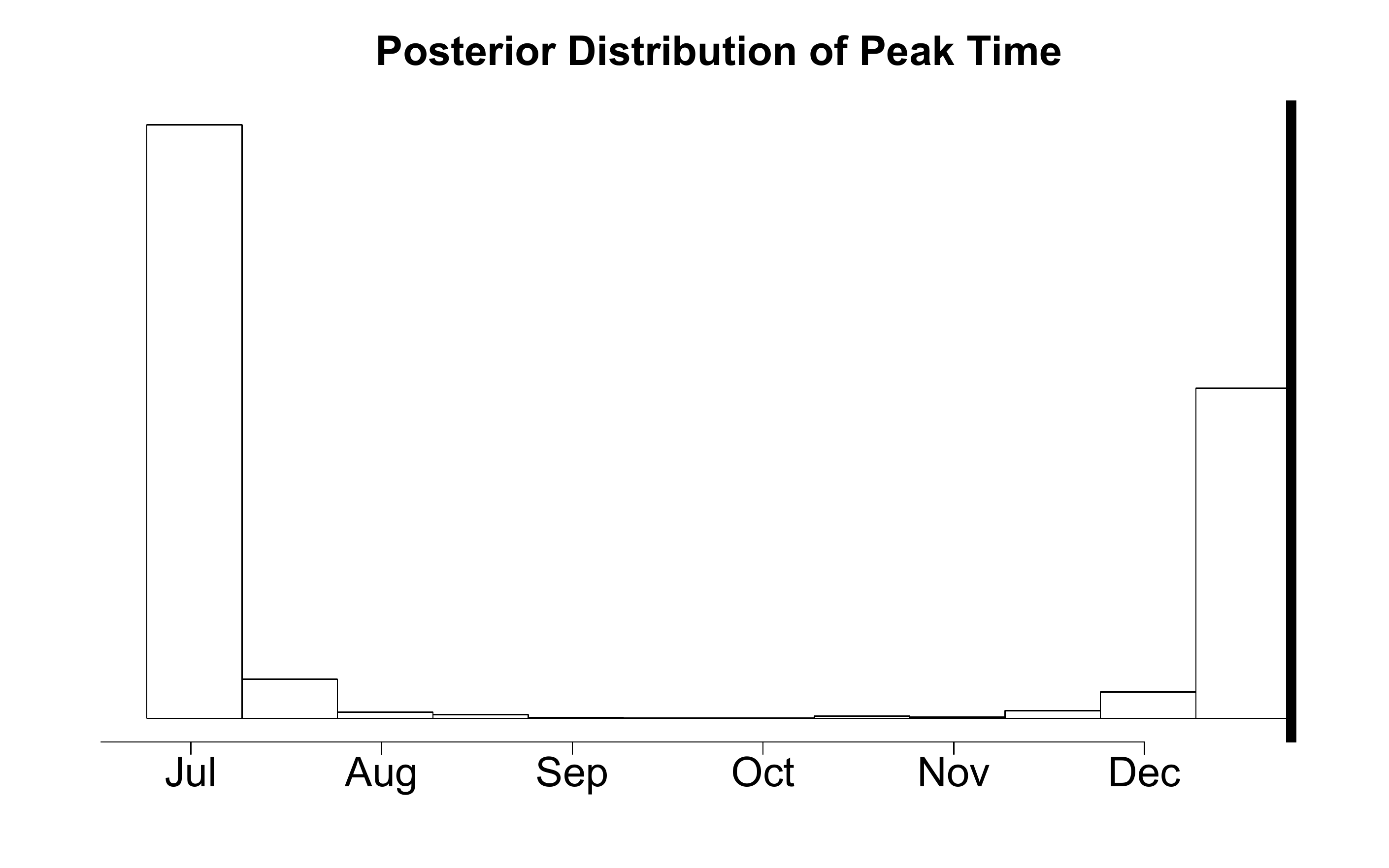}
\includegraphics[width=.4\textwidth]{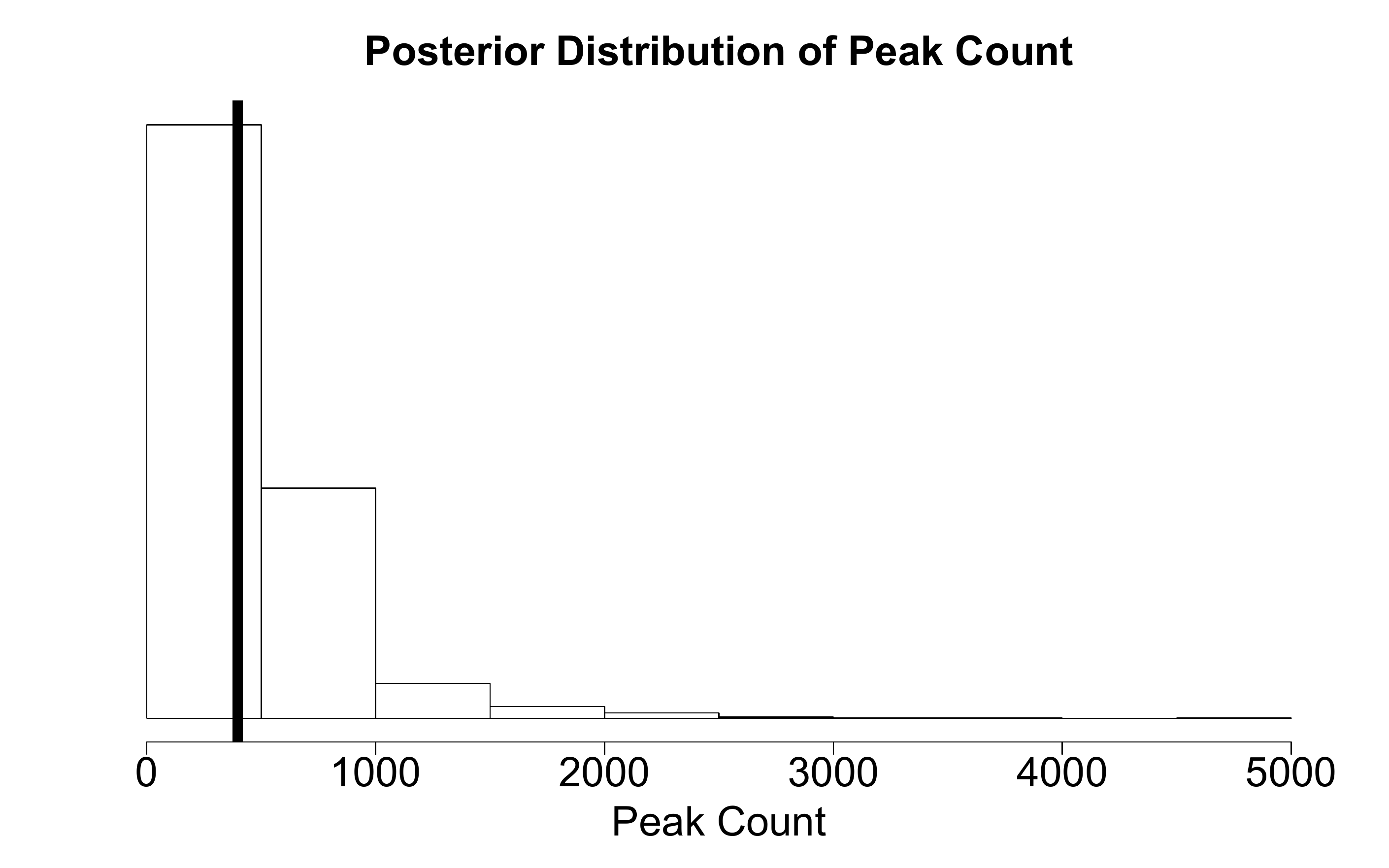}
\caption{An identical format as Figure \ref{fig:emp10-52}, but given observed data up to 26 weeks. The posterior predictive intervals have narrowed considerably. The posterior posterior distribution of peak time is now bimodal, which is accounted for in the disjoint posterior credible region (purple). However, the posterior distribution of peak count remains unimodal, with the bulk of the posterior mass is below 1000 measles counts. 
 \label{fig:emp26-52}}
\end{center}
\end{figure}

An illustration of the forecasting design is given in Figure \ref{fig:emp10-52} for 1961, which immediately precedes the introduction of the measles vaccine. Given observations from the first $m_0=9$ weeks (yellow region), we use the posterior predictive distribution under model  \eqref{nb-model}, \eqref{process}, \eqref{fdlm}, and \eqref{far} to compute (i) the posterior predictive expected value \eqref{cond-mean-2} for the future counts in weeks 10-52 (blue line), (ii) 95\% posterior predictive intervals for the future counts in weeks 10-52 (gray region), and (iii) the posterior predictive distribution for the future peak measles count and the time at which it occurs. This figure appears in color in the electronic version of this article, and color refers to that version. The measles forecasts are accurate, with prediction intervals that widen during peak times from March-July, narrow for July-November, and widen again in December, which matches the observed pattern of variability in Figure \ref{fig:counts}. For both peak value and peak time, the corresponding posterior predictive distribution is centered around the future observed value. An analogous plot given observations from the first $m_0 = 25$ weeks in 1961 is in Web Figure \ref{fig:emp26-52}: the forecasting intervals are far more narrow, yet the posterior distribution of peak time has an interesting bimodality, which assigns large probability to the peak time occurring in  either July or December. 

\subsubsection{Forecasting Methods}
We consider a variety of functional, seasonal, and time series forecasting methods with different distributional assumptions. For each forecasting year $t$, we are interested in computing point and interval forecasts for an unobserved (future) count $Z_t(\tau^*)$ for weeks $\tau^* > m_0$  with known offset $E_t(\tau^*)$ based on (i) historical data from previous years $\{Z_i(\cdot), E_i(\cdot): i < t\}$ and (ii) partial observations from the current year $\{Z_t(\tau): \tau \le m_0 \}$. While accurate point forecasts are undoubtedly important, well-calibrated and precise interval forecasts provide useful information for planning and allocation of resources. In addition to forecasting intervals for future measles counts, we consider 95\% prediction intervals for the \emph{peak count}, $\max \{Z_t(\tau^*): \tau^* > m_0\}$, and \emph{peak time}, $\arg\max_{\tau^*} \{Z_t(\tau^*): \tau^* > m_0\}$, for each year $t$ from 1950-1980. Note that these quantities are not readily available for all methods. 

We consider two variations of the proposed integer-valued functional time series model: a negative binomial distribution with unknown $r$ ({\bf NB-FTS}) and an approximate Poisson distribution ({\bf Pois-FTS}) with $r = 1000$. Forecasts are computed by treating $Z_t(\tau^*)$ as missing data for $\tau^* > m_0$ and imputing from the posterior predictive distribution (see Section \ref{mcmc}). We select $K=6$ and run the MCMC for 30,000 iterations, discarding the first 5,000 iterations as a burn-in and retaining every 5th simulation to reduce autocorrelation. 

As a direct competitor, we include the Gaussian functional time series model ({\bf Gauss-FTS}) of \cite{kowal2018dynamic}, which uses the same model for \eqref{fdlm} and \eqref{far}, but replaces $\theta_i(\tau)$ in \eqref{process} with the observed functional data. Although not well-defined for integer-valued data, Gauss-FTS uses the variance-stabilizing transformation for the Poisson distribution: the input data are $Y_i(\tau) = \sqrt{Z_i(\tau)/E_i(\tau)}$ and the posterior predictive distribution of $E_t(\tau^*) \left[Y_t(\tau^*)\right]^2$ is computed for forecasting and inference. We use the same choice of $K$ and MCMC specifications as for NB-FTS and Pois-FTS. 
Among other functional data methods, we compute forecasts based on (i) the pointwise sample means of the functional observations rescaled by the offsets ({\bf Mean-FDA}), $\hat Z_t(\tau^*) = E_t(\tau^*) \left[\frac{1}{t-1}\sum_{i < t} Z_i(\tau^*)/E_i(\tau^*) \right]$, and (ii)  the previous functional observation rescaled by the offsets ({\bf RW-FDA}), $\hat Z_t(\tau^*) = E_t(\tau^*)\left[Z_{t-1}(\tau^*)/E_{t-1}(\tau^*) \right]$, i.e., a random walk forecast. While these are nominally functional data estimators, they are also seasonal time series estimators, and therefore provide an important baseline. 


We also include more classical seasonal and integer-valued time series models, which are broadly popular for disease forecasting \citep{shumway2000time,martinez2011sarima}. Using the time series data $z_1,\ldots, z_{m_0 + (t-1)m}$ up to week $m_0$ of the current year $t$, we fit a seasonal autoregressive moving average ({\bf SARIMA}) model using the \texttt{auto.arima()} function in the \texttt{forecast} package in \texttt{R} \citep{forecastR,forecastVig} with the order of the autoregressive, moving average, seasonal autoregression, and seasonal moving average components (based on $m=52$ week seasonality) selected by AIC. Similarly, we fit a SARIMA model to the square-root transformed count data ({\bf sqrt-SARIMA}) to stabilize the variance. In both cases, the counts are scaled by the offset before fitting and rescaled for forecasting and inference. Among time series models for count data, we include negative-binomial ({\bf NB-TS}) and Poisson ({\bf Pois-TS}) models implemented using the \texttt{tsglm()} function in the \texttt{tscount} package in \texttt{R} \citep{tscountR,tscountVig}. We select a log-link function and include past observations and past means from lags 1,2, 52, and 53, which incorporates local time-dependence and annual seasonality. 

\subsubsection{Forecasting Evaluation Metrics}
Forecasting performance is evaluated based on (i) point forecasts, (ii) interval forecasts, and (iii) interval forecasts of epi-relevant features, namely the peak count and the peak time. In each case, results are divided into pre- and post-vaccine eras, with 14 years pre-vaccine and 16 years post-vaccine (we omit 1964 due to substantial missingness). For each year, we forecast $m - m_0$ weeks of measles counts, totaling 602 weeks pre-vaccine and 688 weeks post-vaccine for $m_0 = 9$, and 378 weeks pre-vaccine and 432 weeks post-vaccine for $m_0 = 25$. 

For a point forecast, say $\hat Z_i(\tau^*)$, we consider two definitions of mean absolute error (MAE). We compute the MAE averaged across all weeks $\tau^* > m_0$ in each year $i$,  $\mbox{MAE}_i = (m - m_0)^{-1}\sum_{\tau^* = m_0 + 1}^{m} | Z_i(\tau^*) - \hat Z_i(\tau^*)|$, which illustrates how forecasting performance varies from year-to-year. To compare forecasting performance for different forecasting horizons, we also compute the MAE averaged across all years in each week $\tau^*$, $\mbox{MAE}(\tau^*) = (n-n_0)^{-1}\sum_{i = n_0 + 1}^{n} | Z_i(\tau^*) - \hat Z_i(\tau^*)|$ where $n_0$ is the number of years used for estimation and $n$ is the total number of years. 

Interval forecasts are evaluated using empirical coverage probability (ECP) and median interval width (MIW): for a prediction interval $(\hat Z_i^{lower}(\tau^*), \hat Z_i^{upper}(\tau^*))$, $\mbox{ECP} = (n  -n_0)^{-1}(m - m_0)^{-1} \sum_{i=n_0 + 1}^n \sum_{\tau^* = m_0 + 1}^m \mathbb{I}\Big\{\hat Z_i^{lower}(\tau^*) \le Z_i(\tau^*) \le \hat Z_i^{upper}(\tau^*)\Big\}$  and $\mbox{MIW} = \mbox{median}\{\hat Z_i^{upper}(\tau^*) - \hat Z_i^{lower}(\tau^*): i=n_0+1,\ldots, n, \tau^* = m_0+1,\ldots,m\}$. Both metrics are important: ECP assesses calibration of the interval, while MIW describes the precision of the uncertainty quantification. The goal is to obtain the most narrow forecasting intervals that achieve the 95\% nominal coverage. 

\subsubsection{Forecasting Results}
\begin{figure}[h]
\begin{center}
\includegraphics[width=.49\textwidth]{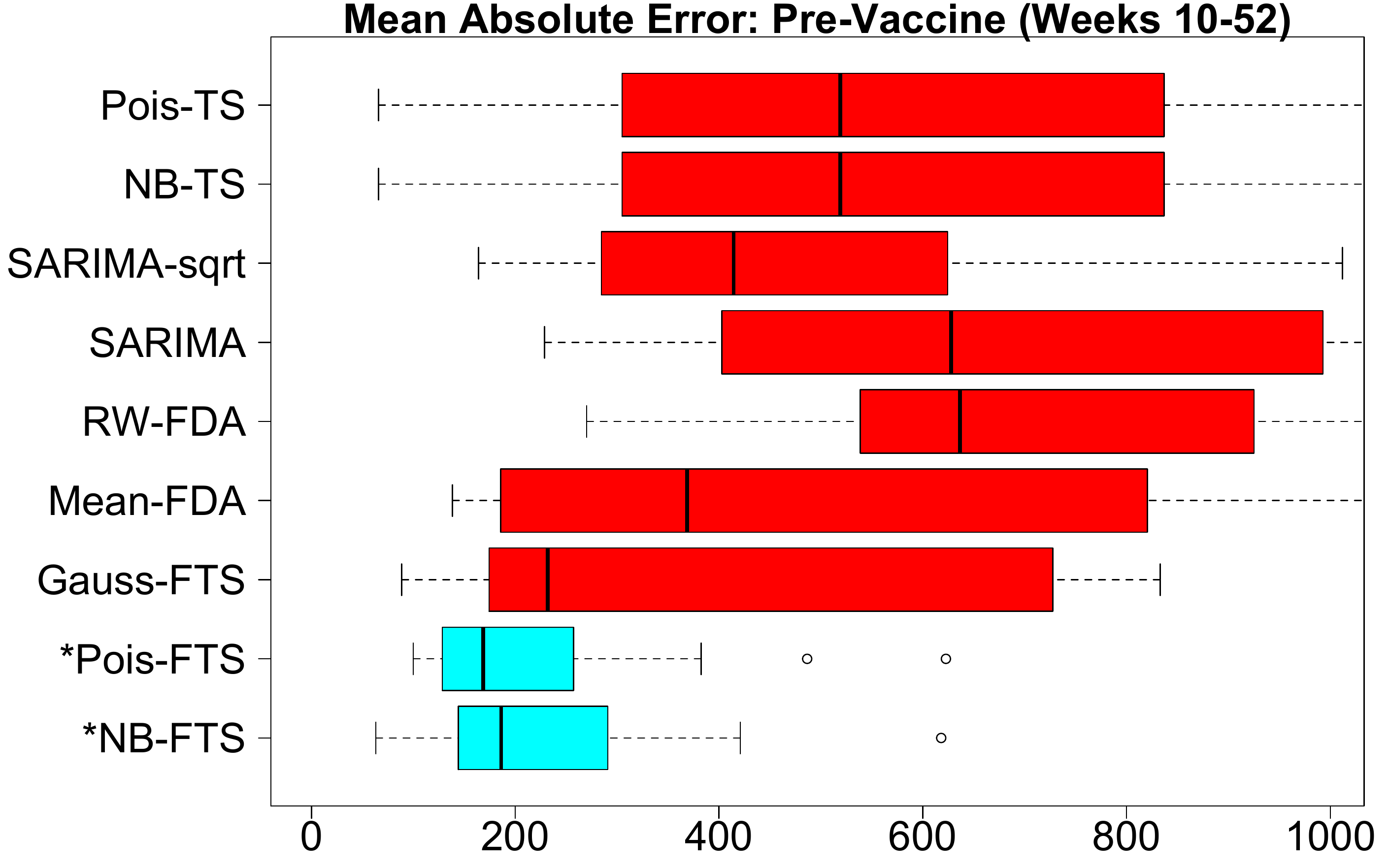}
\includegraphics[width=.49\textwidth]{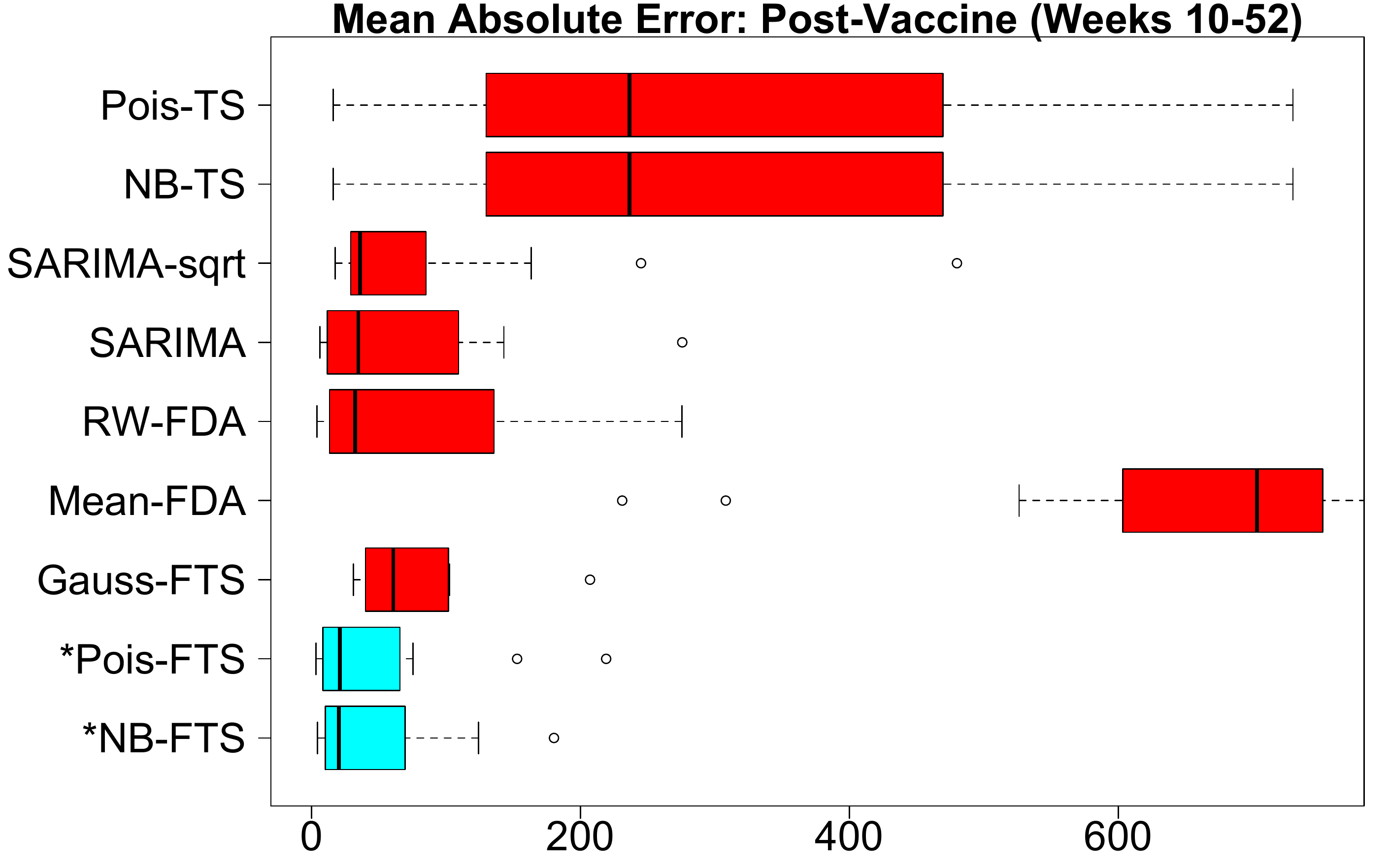}

\includegraphics[width=.49\textwidth]{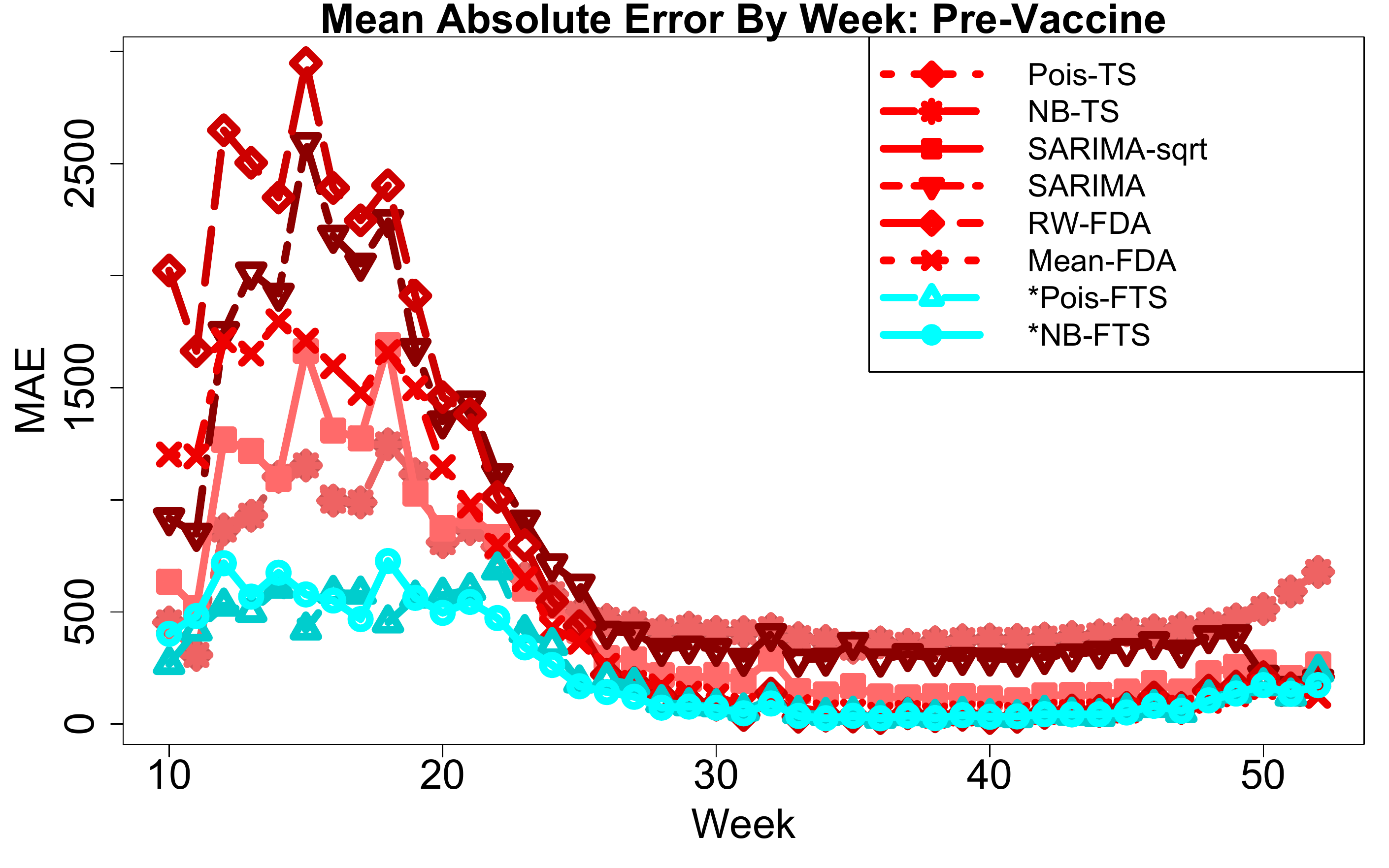}
\includegraphics[width=.49\textwidth]{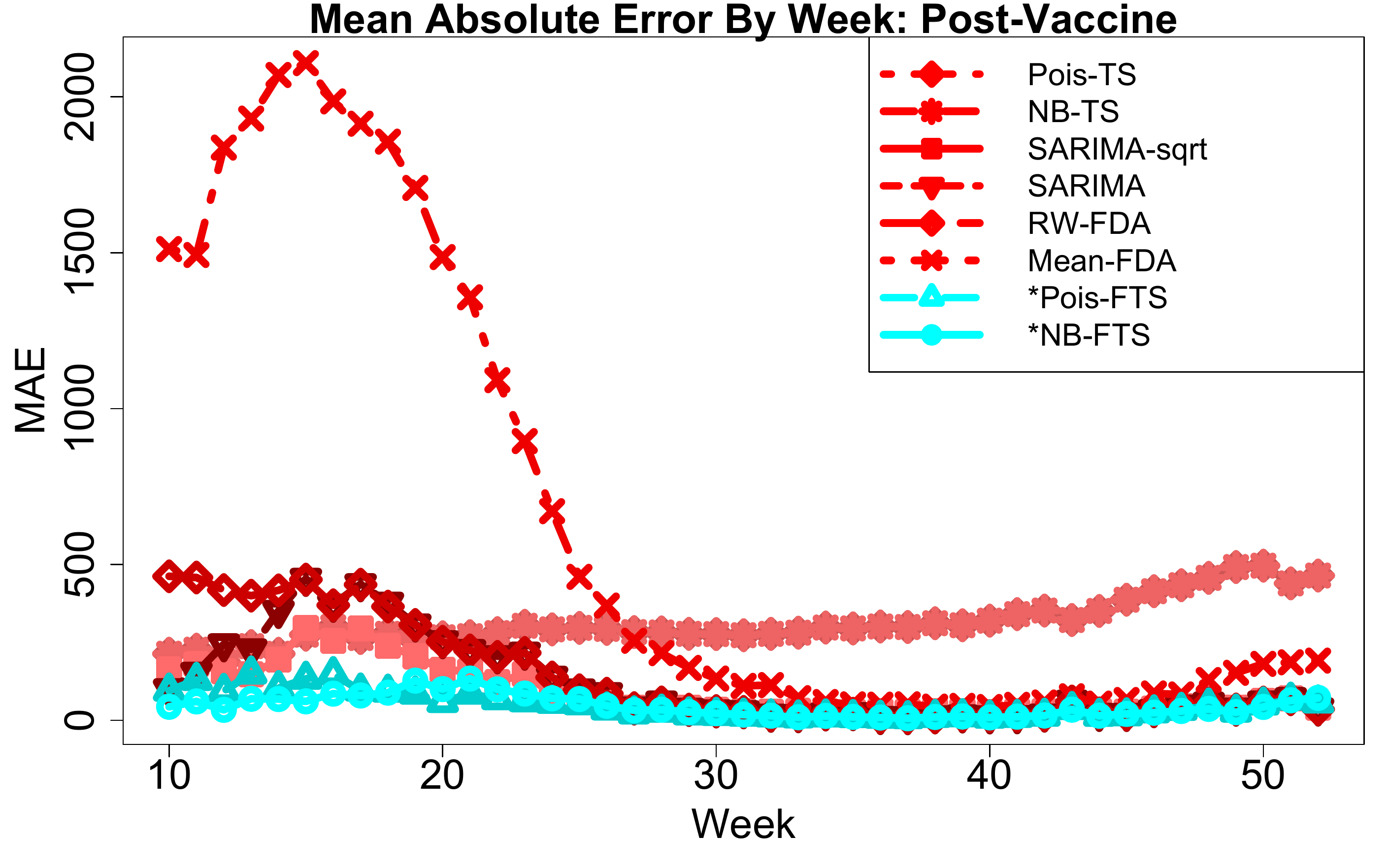}

\caption{Mean absolute errors pre-vaccine ({\bf left}) and post-vaccine ({\bf right}) for forecasting weeks 10-52.  {\bf Top:} $\mbox{MAE}_i$ across years $i$ and aggregated by week. {\bf Bottom:} $\mbox{MAE}(\tau^*)$ for each week $\tau^*$ and aggregated by year. Forecasting is most difficult pre-vaccine, particularly in the first half of the year. The proposed methods *NB-FTS and *Pois-FTS (cyan) perform best in all settings---across years, across weeks-ahead forecasts, and pre-and post-vaccine---with the largest gains pre-vaccine during the peak weeks 10-20. Extreme outliers from competing methods were omitted for display purposes, including Gauss-FTS in the bottom plots. This figure appears in color in the electronic version of this article, and color refers to that version. 
 \label{fig:mae10-52}}
\end{center}
\end{figure}

\begin{figure}[h]
\begin{center}
\includegraphics[width=.49\textwidth]{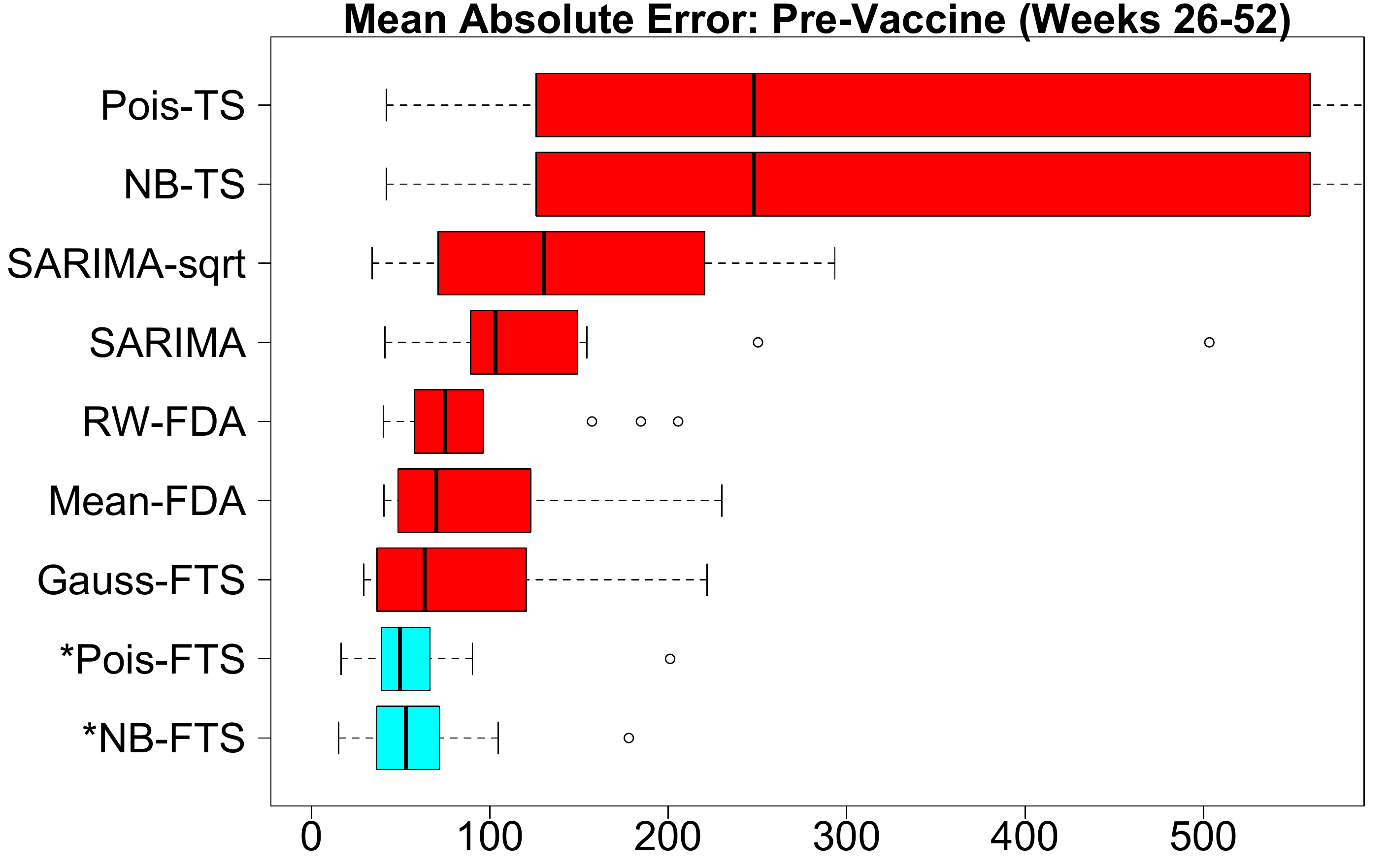}
\includegraphics[width=.49\textwidth]{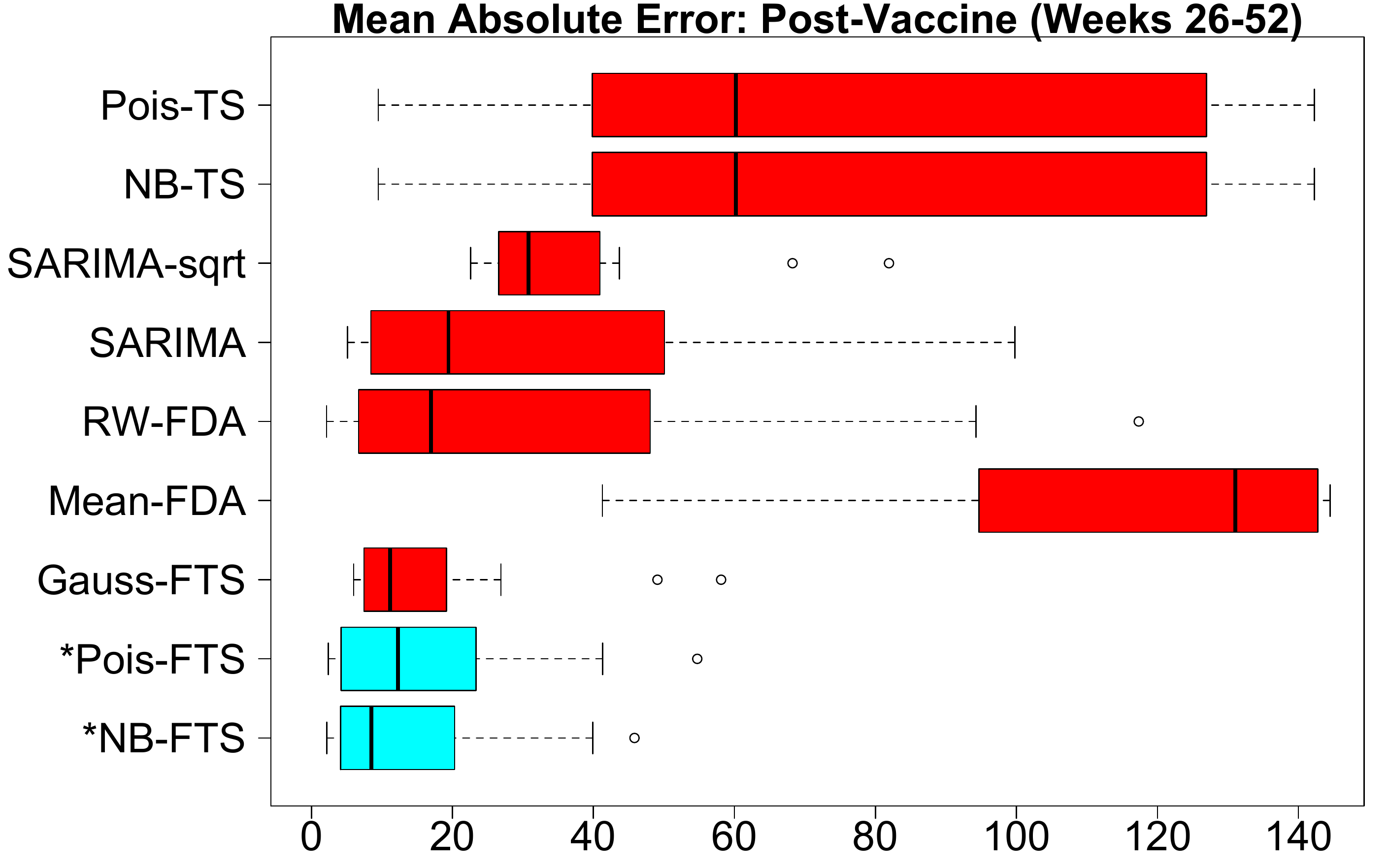}

\includegraphics[width=.49\textwidth]{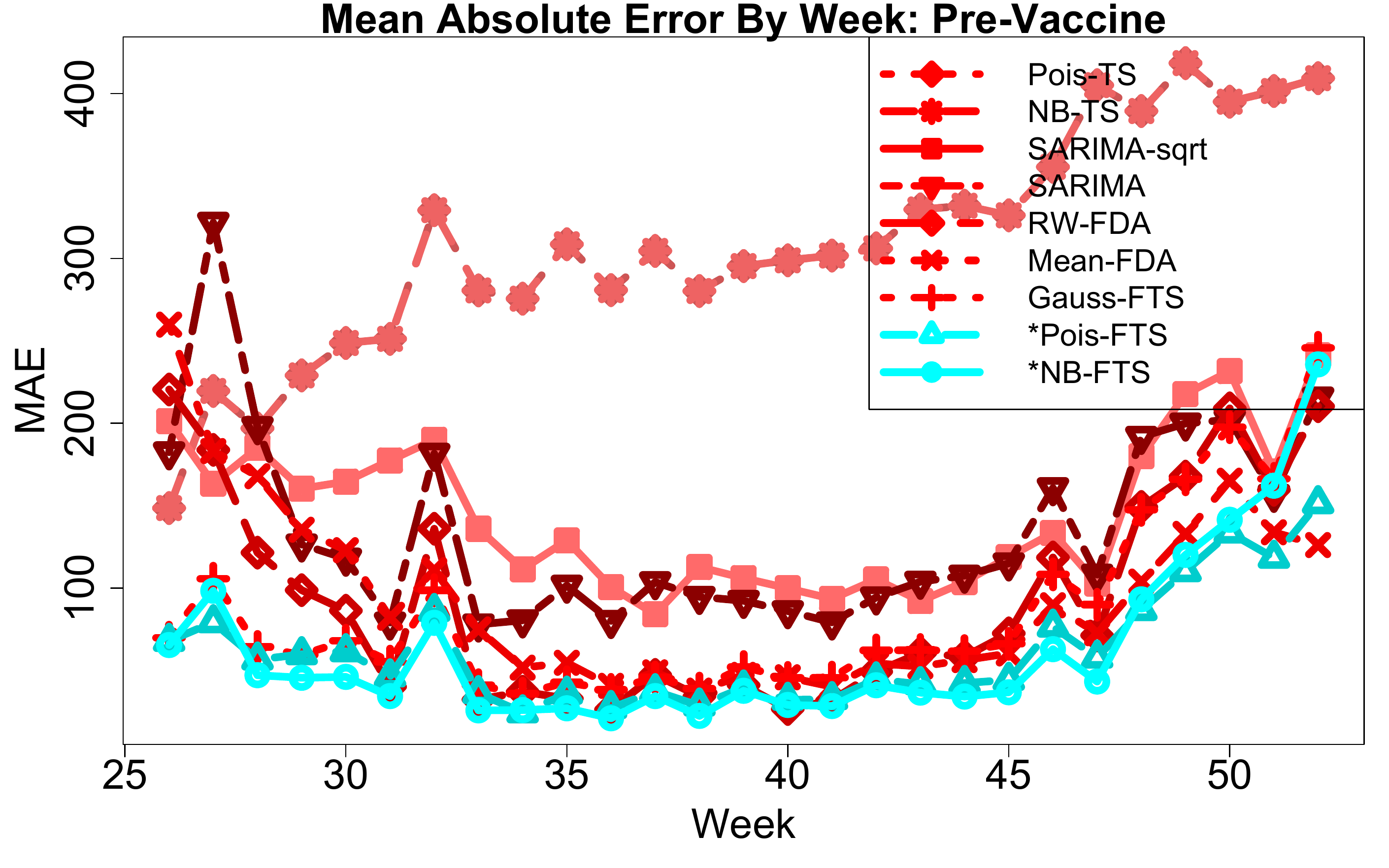}
\includegraphics[width=.49\textwidth]{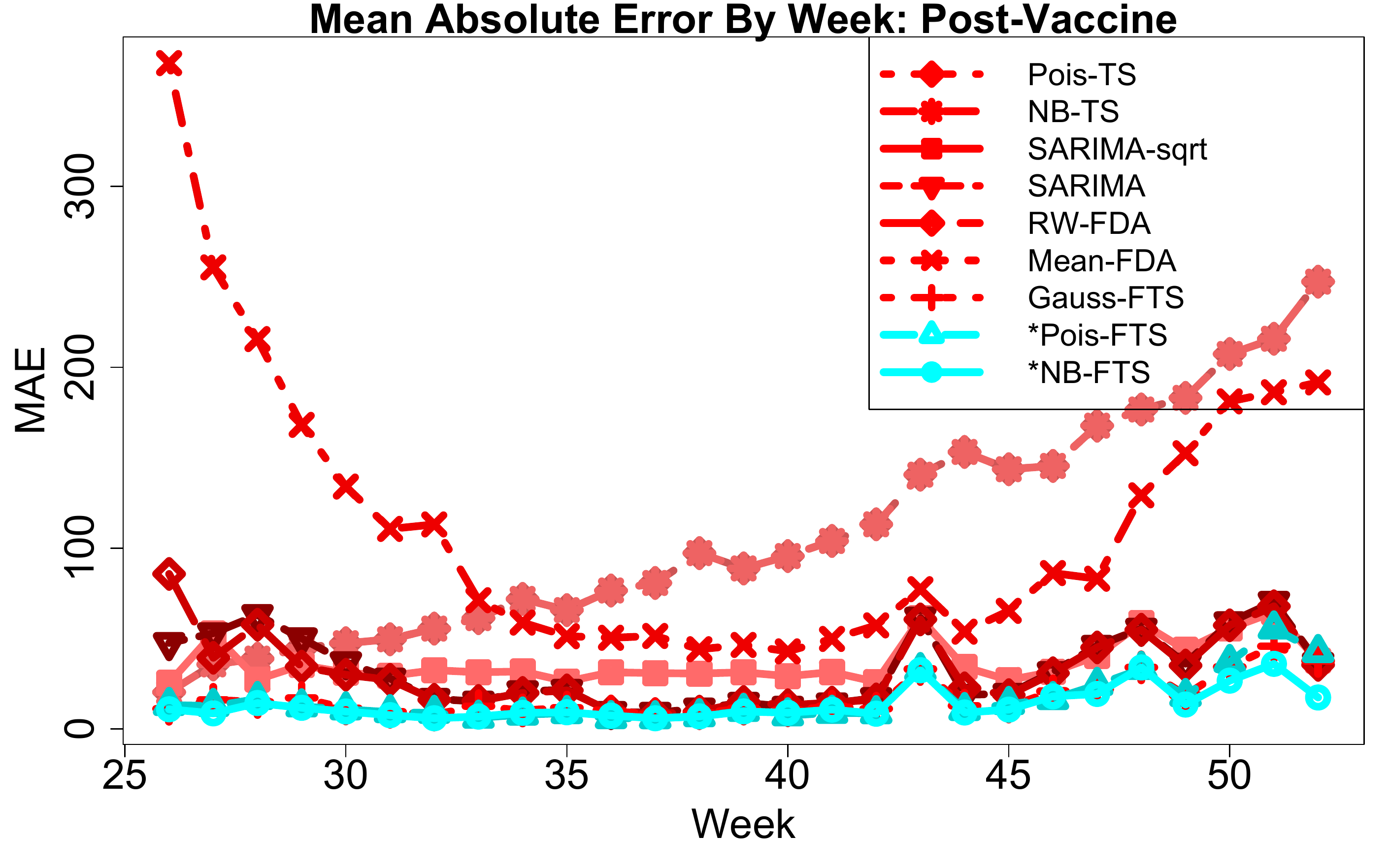}
\caption{Mean absolute errors pre-vaccine ({\bf left}) and post-vaccine ({\bf right}) for forecasting weeks 26-52.  {\bf Top:} $\mbox{MAE}_i$ across years $i$ and aggregated by week. {\bf Bottom:} $\mbox{MAE}(\tau^*)$ for each week $\tau^*$ and aggregated by year. Forecasting is most difficult pre-vaccine, particularly in the first half of the year. The proposed methods *NB-FTS and *Pois-FTS (cyan) perform best in all settings--across years, across weeks-ahead forecasts, and pre-and post-vaccine. Extreme outliers from competing methods were omitted for display purposes.
 \label{fig:mae26-52}}
\end{center}
\end{figure}

A comparison of the point forecasts is given in Figure \ref{fig:mae10-52}, which includes both MAE metrics for forecasting weeks 10-52. The proposed methods NB-FTS and Pois-FTS perform the best across all years, with notable improvements over competing methods in the pre-vaccine era. The specific time-of-year improvements are illustrated in the $\mbox{MAE}(\tau^*)$ plots in the bottom panels: the NB-FTS and Pois-FTS forecasts are substantially more accurate during weeks 10-20, which corresponds to peak measles counts, especially pre-vaccine. Unsurprisingly, Mean-FDA performs worst post-vaccine, since the forecasts are based on pre- and post-vaccine averages. The results for weeks 26-52 are in Web Figure \ref{fig:mae26-52}: the proposed methods NB-FTS and NB-Poisson remain superior, albeit by smaller margins. 

\begin{figure}[h]
\begin{center}
\includegraphics[width=.49\textwidth]{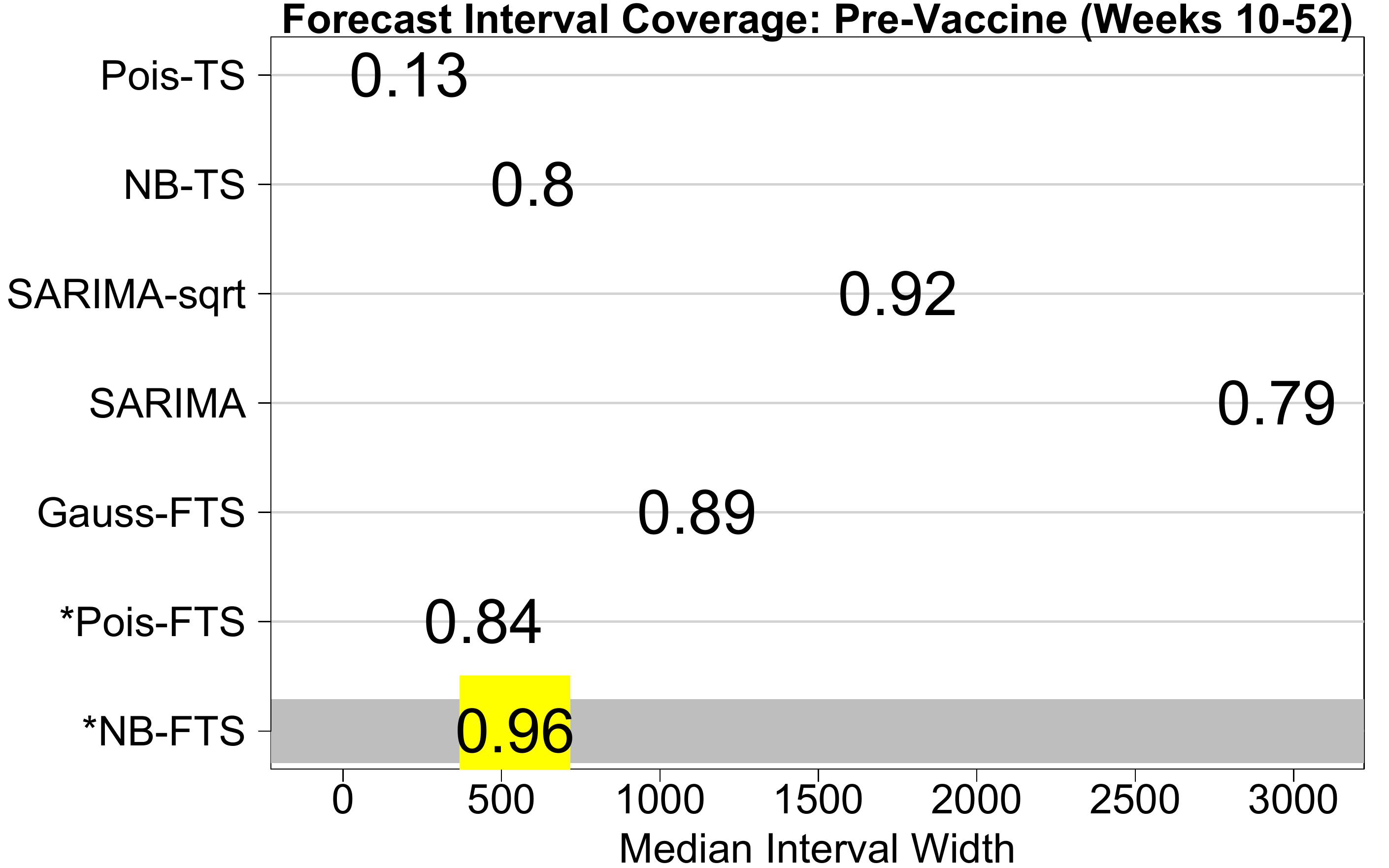}
\includegraphics[width=.49\textwidth]{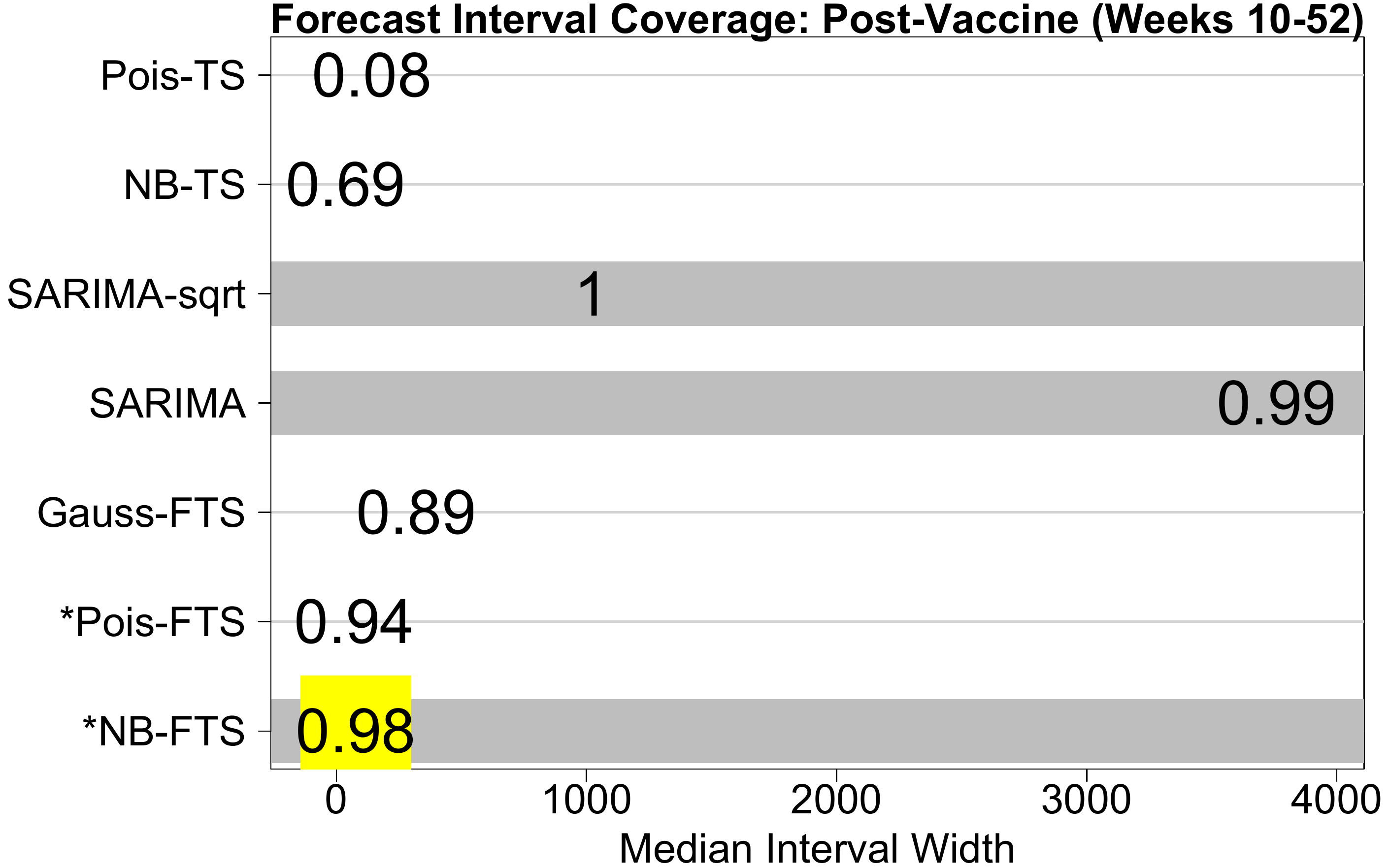}
\caption{Median forecast interval width pre-vaccine ({\bf left}) and post-vaccine ({\bf right}) for weeks 10-52, with labels indicating the empirical coverage probability. Methods achieving the 95\% nominal coverage are banded in grey; among these methods, the one with the narrowest intervals is highlighted in yellow. The proposed *NB-FTS provides the correct coverage with substantially more narrow intervals than computing methods.  This figure appears in color in the electronic version of this article, and color refers to that version. \label{fig:cover10-52}}
\end{center}
\end{figure}

\begin{figure}[h]
\begin{center}
\includegraphics[width=.49\textwidth]{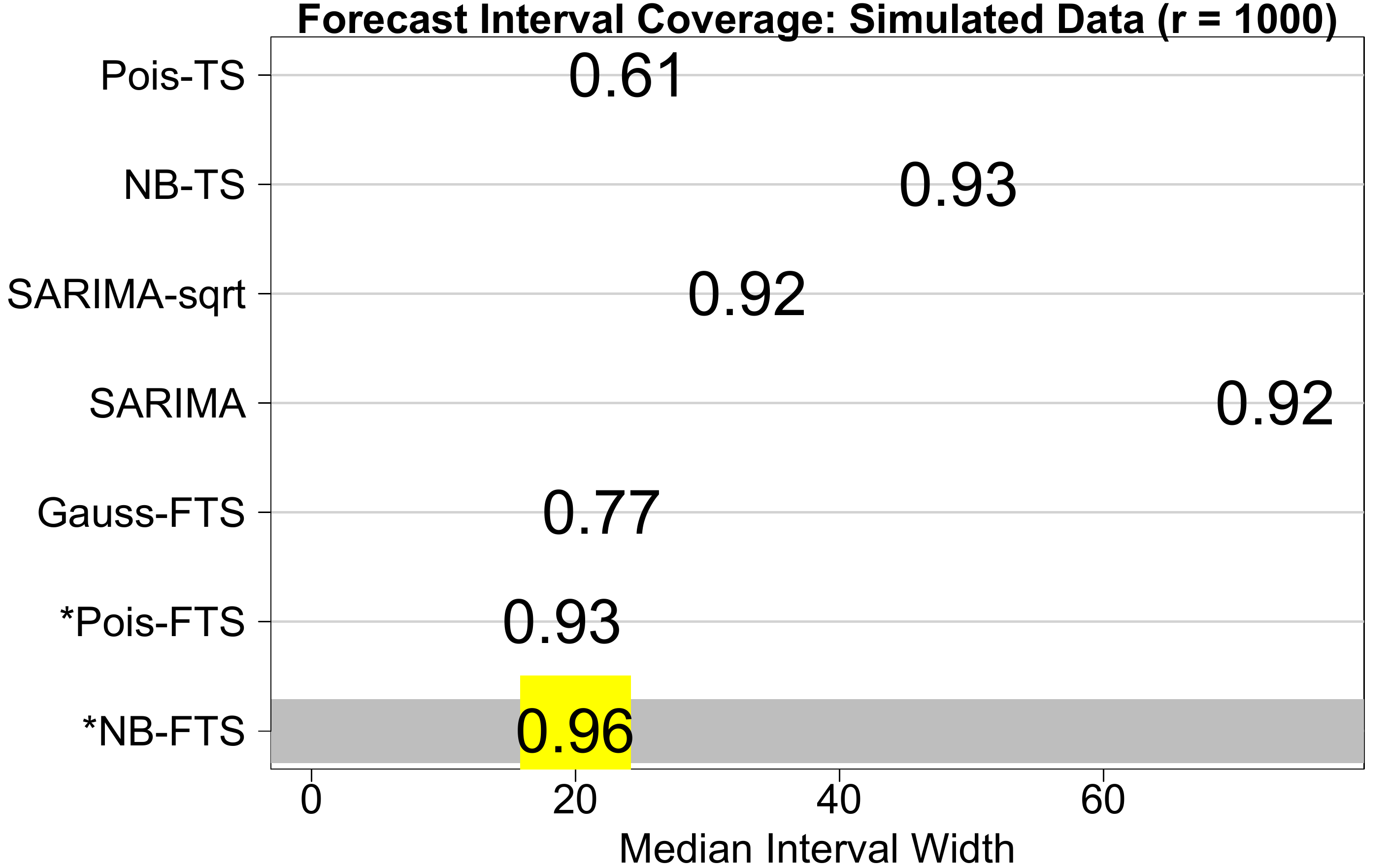}
\includegraphics[width=.49\textwidth]{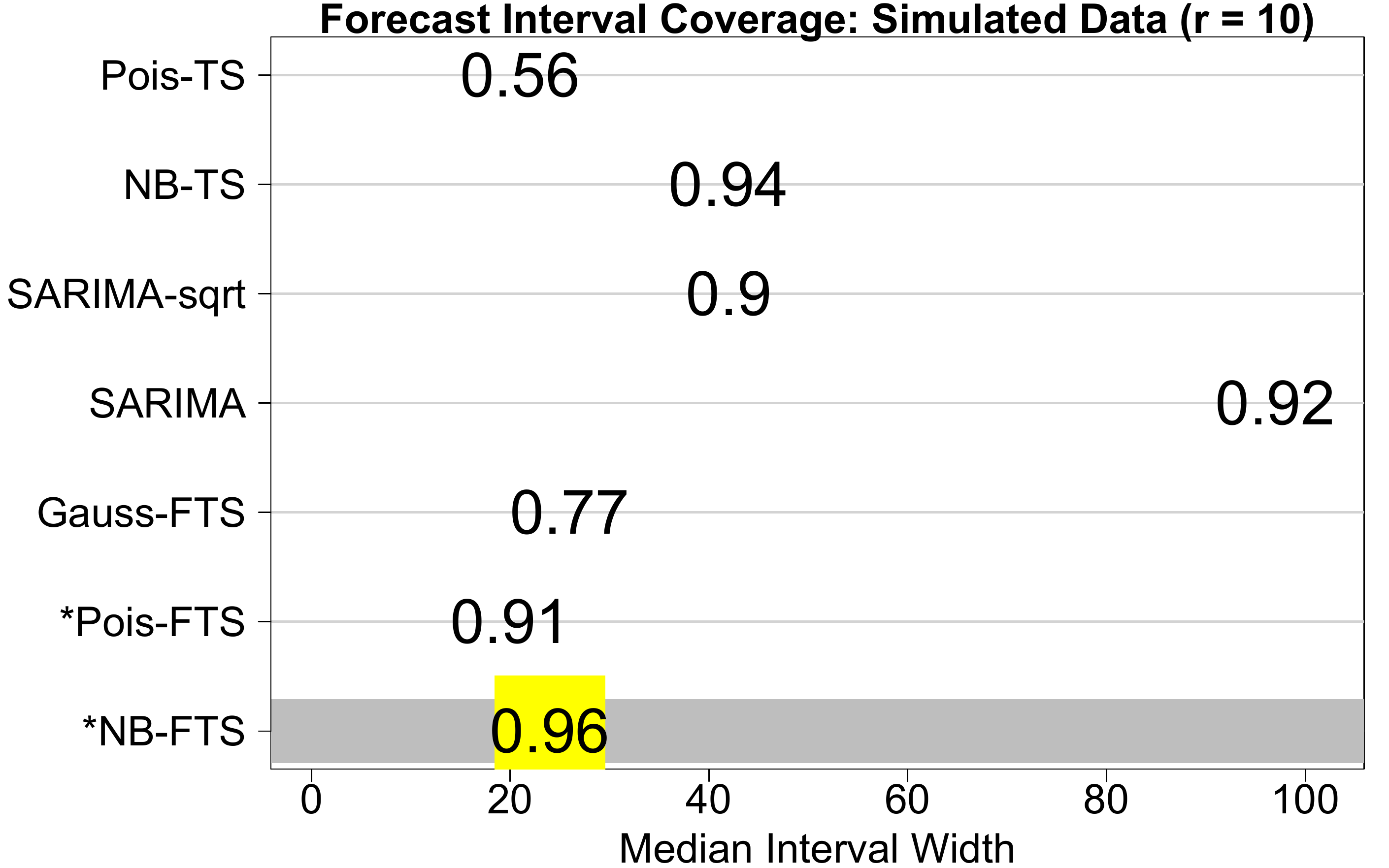}
\caption{Median forecast interval width for $r=1000$ ({\bf left}) and $r=10$ ({\bf right}) for forecasting the remaining time points 31-50, with labels indicating the empirical coverage probability. Methods achieving the 95\% nominal coverage are banded in grey; among these methods, the one with the narrowest intervals is highlighted in yellow. This figure appears in color in the electronic version of this article, and color refers to that version. \label{fig:cover-sim}}
\end{center}
\end{figure}

\begin{figure}[h]
\begin{center}
\includegraphics[width=.49\textwidth]{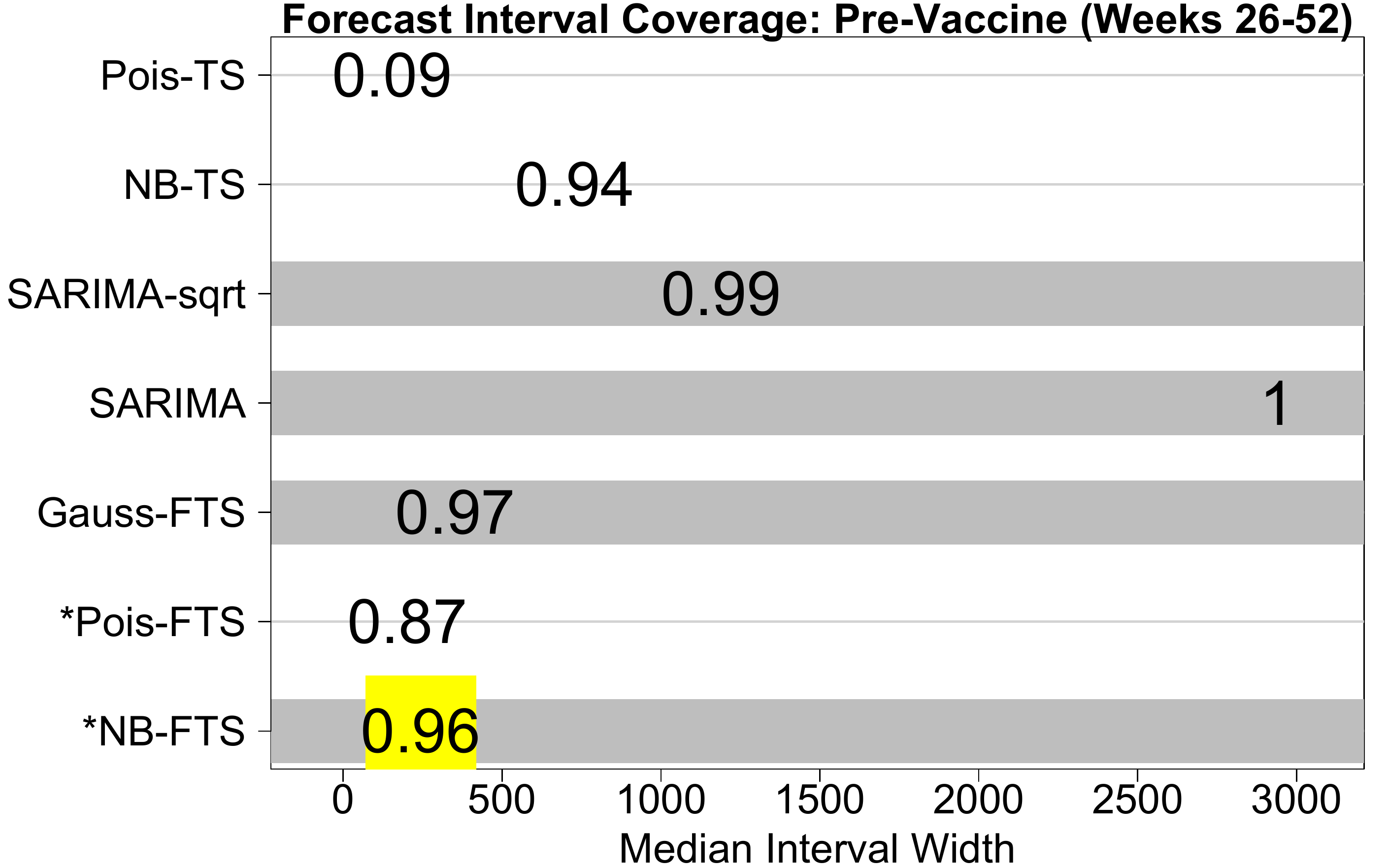}
\includegraphics[width=.49\textwidth]{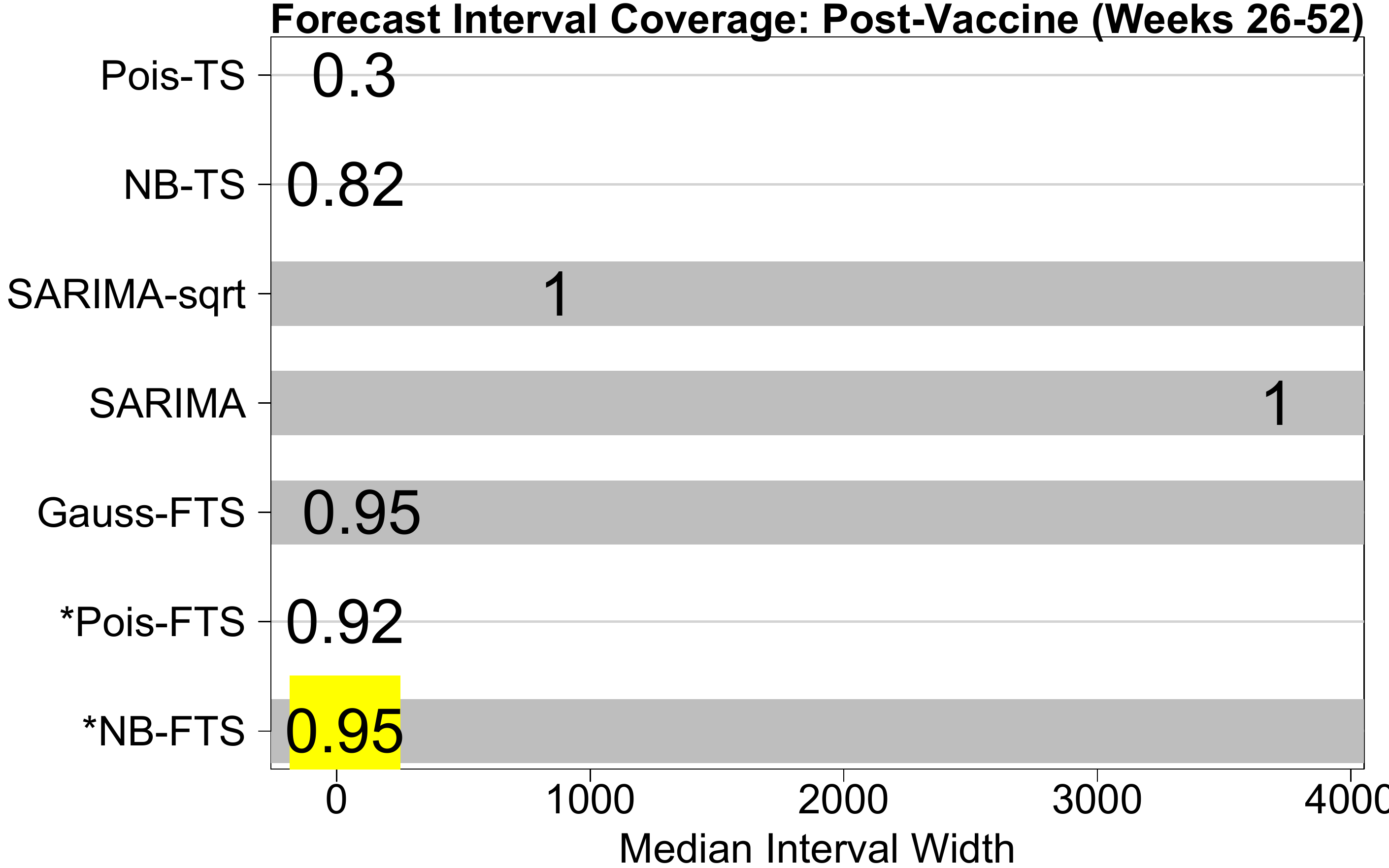}
\caption{Median forecast interval width pre-vaccine ({\bf left}) and post-vaccine ({\bf right}) for forecasting weeks 26-52, with labels indicating the empirical coverage probability. Methods achieving the 95\% nominal coverage are banded in grey; among these methods, the one with the narrowest intervals is highlighted in yellow. \label{fig:cover26-52}}
\end{center}
\end{figure}

The most striking results are the forecasting interval comparisons for weeks 10-52 (Figure \ref{fig:cover10-52}), weeks 26-52 (Web Figure \ref{fig:cover26-52}), as well as simulated data from Section \ref{sims} (Web Figure \ref{fig:cover-sim}), with exact ECP and MIW values given in Web Table \ref{table:cover}. Notably, only the proposed NB-FTS model provides forecasting intervals that consistently achieve the 95\% nominal coverage. Naturally, correct coverage is essential for utility and interpretability of the intervals. In addition to correct coverage, NB-FTS produces substantially more narrow forecasting intervals, especially among methods that achieve nearly the nominal 95\% coverage, which implies greater precision in the forecasting intervals. For example, in the pre-vaccine era, NB-FTS forecast intervals achieve 96\% coverage with a median width of 543 counts, while the most competitive alternative, sqrt-SARIMA,  has 92\% coverage with a median width of 1751 counts---which is more than three times larger. Pois-FTS is competitive with NB-FTS, but typically suffers from undercoverage. Interestingly, the integer-valued distribution appears to be important: Gauss-FTS, despite including a variance-stabilizing transformation, either fails to provide adequate coverage or produces much wider forecast intervals than NB-FTS. 
Clearly, NB-FTS provides superior forecast intervals, which are both correctly calibrated at 95\% and more precise than competing methods.


Similarly, Table \ref{table:cover-epi} compares the 95\% forecasting intervals for the epi-relevant features corresponding to the peak time and peak values of the measles counts over the weeks 10-52, as well as for the simulated data in Section \ref{sims}. For resource allocation and planning, it is important to know when the measles season will peak, and how many people will be affected at the peak. For the Bayesian (integer-valued and Gaussian) functional data models, inference for these quantities are readily available via the posterior predictive distribution (as in Figure \ref{fig:emp10-52}). Impressively, the proposed integer-valued functional data models provide the correct nominal coverage with much narrower intervals than the Gaussian model. 

These results cumulatively demonstrate the clear advantages of using the proposed \emph{integer-valued functional data model} for point, interval, and epi-relevant forecasts of measles counts. The \emph{integer-valued} distribution offers substantial improvements for NB-FTS and Pois-FTS relative to Gauss-FTS, including both point estimation and interval coverage, while the \emph{functional data} approach outperforms seasonal and integer-valued time series models, especially among forecast intervals and pre-vaccine point forecasts.

\subsection{Measles Counts: Inference}\label{measles:inf}
The proposed modeling framework may be used for in-sample estimation,  inference, and imputation of missing values. Consider the introduction of the measles vaccine in 1963 in Figure \ref{fig:counts}: while there is a notable change in the measles incidence pattern, the decline in measles counts is non-monotone due to the presence of seasonality. A functional time series approach offers an intuitive decomposition: the curves $f_k(\tau)$ model intra-year seasonality, while the basis coefficients $\beta_{k,i}$ capture how the level and seasonality vary from year-to-year. We propose the following extension of \eqref{far} to include predictors $\{x_{i,j}\}_{j=1}^p$ in year $i$, such as yearly time trends and a vaccination effect:
\begin{equation}\label{far-reg}
\beta_{k,i} = \mu_{k} +  \sum_{j=1}^p x_{i,j} \alpha_{j,k} + \gamma_{k,i}, \quad 
\gamma_{k,i} = \phi_k \gamma_{k,i-1} + \eta_{k,i} 
\end{equation}
with $\eta_{k,i}$ distributed as in \eqref{far}, which is a multiple linear regression with autoregressive errors for each basis coefficient $k=1,\ldots,K$. Model \eqref{far-reg} is utilized in \cite{kowal2018dynamic} for Gaussian function-on-scalars regression, but is easily adaptable to the integer-valued setting via the MCMC algorithm in Section \ref{mcmc}.  For interpretability, we may rewrite the conditional mean of the count observations \eqref{cond-mean-2} as follows:
\begin{equation}\label{cond-mean-3}
\mathbb{E}[Z_i(\tau) |  \mu_i(\cdot), \sigma_\epsilon, r] = \exp\left[\tilde \mu(\tau)\right] \left\{\prod_{j=1}^p \exp\left[x_{i,j} \tilde \alpha_j(\tau)\right] \right\}
\exp\left[\tilde \gamma_i(\tau)\right] \exp(\sigma_e^2/2) 
 \end{equation}
where $\tilde \mu(\tau) = \sum_k f_k(\tau)\mu_k$ is an intercept function, $\tilde \alpha_j(\tau) = \sum_k f_k(\tau) \alpha_{j,k}$ is the regression function for predictor $j$, and $\tilde \gamma_i(\tau) = \sum_k f_k(\tau) \gamma_{k,i}$ is the year-specific autoregressive random effect. The representation in \eqref{cond-mean-3} is analogous to standard log-linear models, in particular for Poisson and negative-binomial distributions with log-link functions. The seasonal effect of predictor $j$ is capture by $\tilde \alpha_j(\tau)$; when $\tilde \alpha_j(\tau) = 0$ for all weeks $\tau$, the effect is null. Following \cite{kowal2018dynamic}, we assume nested horseshoe priors on the regression coefficients: $\alpha_{j,k} \stackrel{indep}{\sim} N(0, \sigma_{\alpha_{j,k}}^2)$ with $\sigma_{\alpha_{j,k}} \stackrel{indep}{\sim} C^+(0, \lambda_j)$, $\lambda_j \stackrel{iid}{\sim}C^+(0, \lambda_0)$, and $\lambda_0 \sim C^+(0, 1/\sqrt{p})$, which provides a hierarchy of local $(j,k)$ shrinkage, predictor-specific $j$ shrinkage, and global shrinkage.

We implement model \eqref{nb-model}, \eqref{process}, \eqref{fdlm}, and \eqref{far-reg} for the entire measles counts dataset from 1928-2002. For predictors, we include a linear time trend $i$, an indicator of post-vaccine years $\mathbb{I}\{i \ge 1963\}$, and a vaccine-year interaction $i\mathbb{I}\{i \ge 1963\}$. We select $K=6$ and run the MCMC for 30,000 iterations, discarding the first 5,000 iterations as a burn-in and retaining every 5th simulation. Traceplots demonstrate good mixing and suggest convergence, and effective samples sizes are sufficiently large. Missing values are automatically imputed within the Gibbs sampler (see Web Figure \ref{fig:impute}). The model clearly recognizes overdispersion: the 95\% highest posterior density (HPD) interval for the dispersion parameter $r$ is $[4.72, 6.92]$. In addition, the year-to-year dynamics---even after adjusting for predictors---are substantial: posterior credible intervals for autoregressive coefficients $\phi_k$ in \eqref{far-reg} exclude zero for $k=1$ (95\% HPD interval $[0.45,  0.89]$) and $k=2$ (95\% HPD interval $[-0.527, -0.003]$). 

\begin{figure}[h]
\begin{center}
\includegraphics[width=0.32\textwidth]{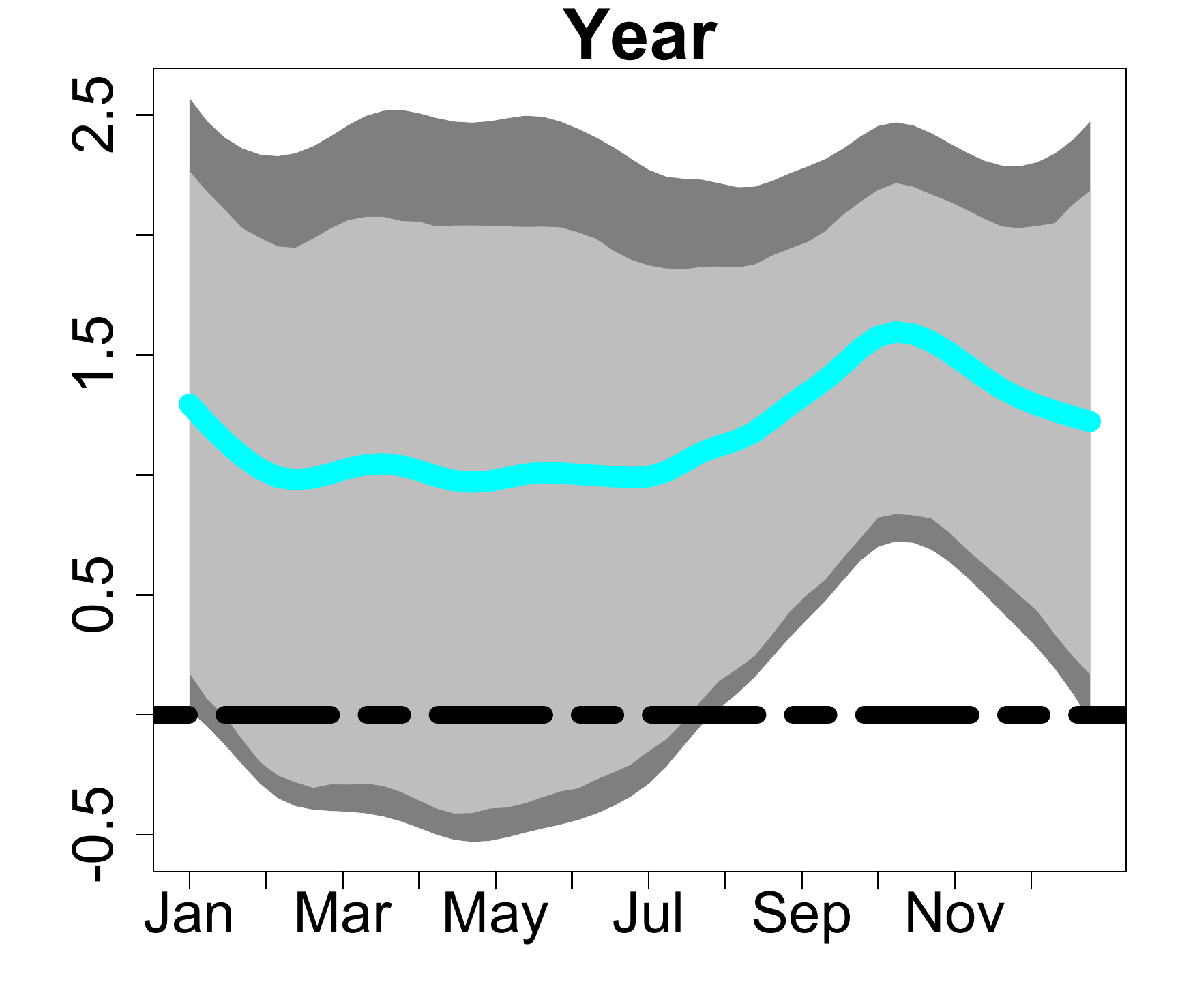}
\includegraphics[width=0.32\textwidth]{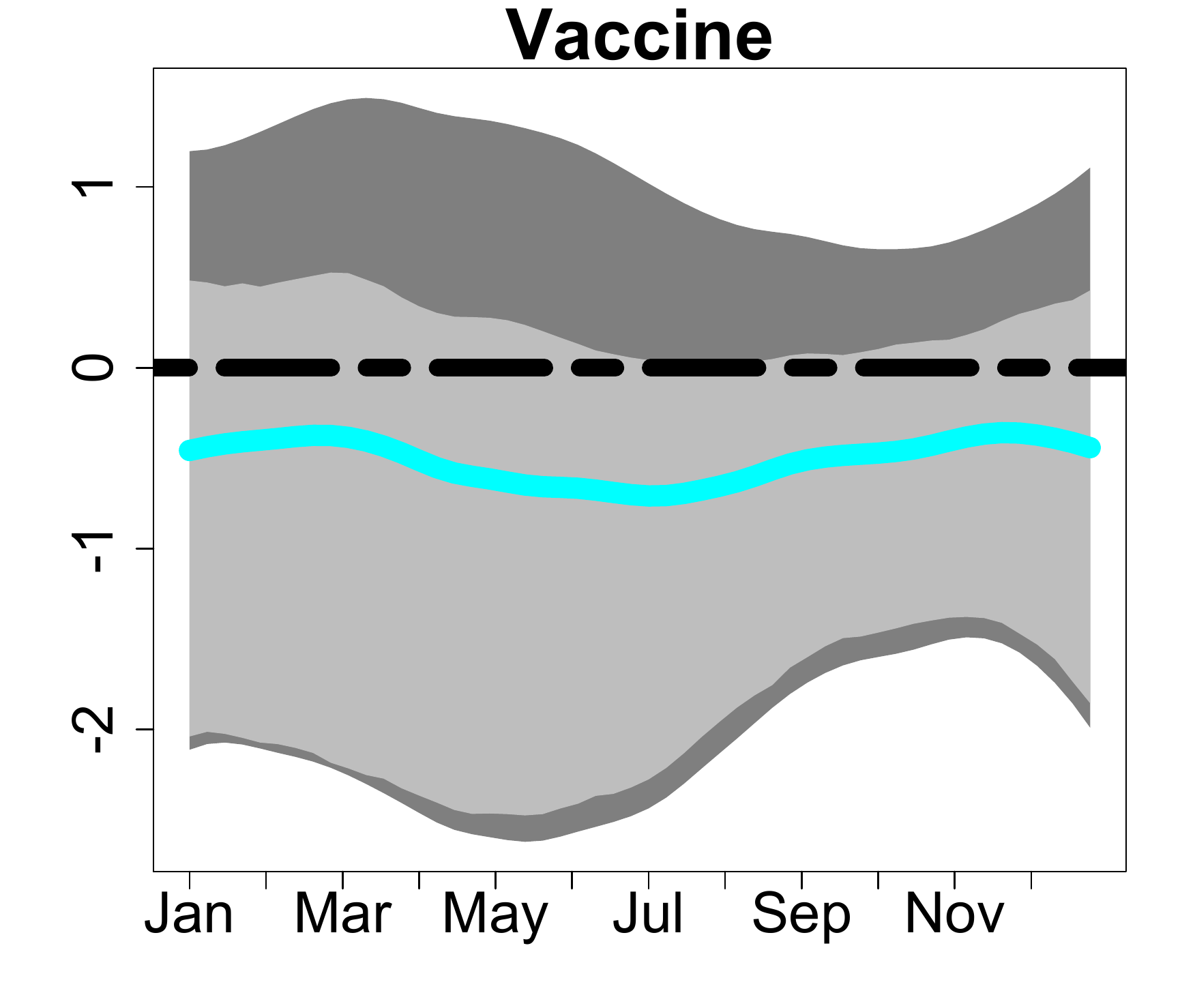}
\includegraphics[width=0.32\textwidth]{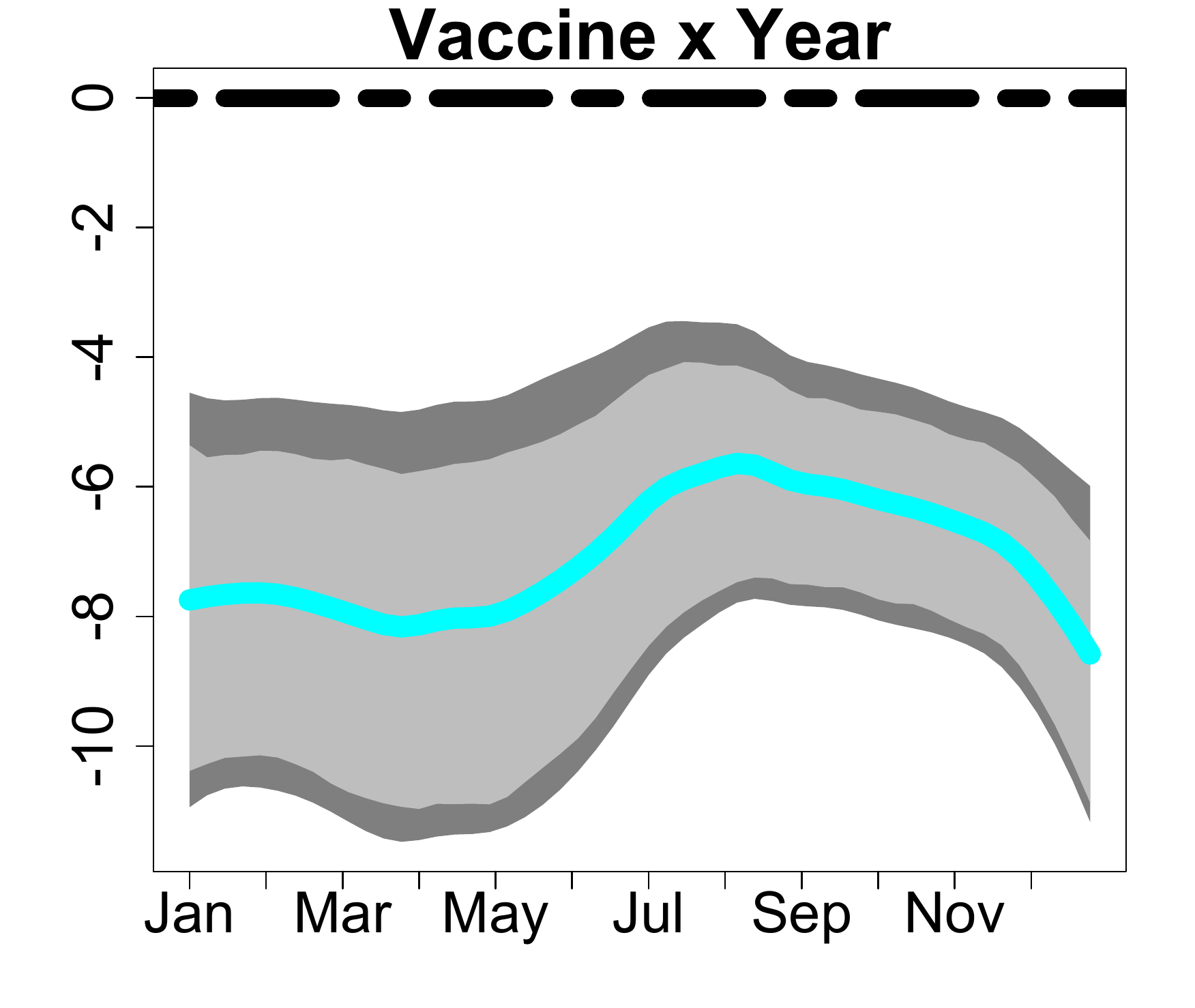}
\caption{Regression coefficient functions $\tilde \alpha_j(\tau)$ for year ({\bf left}), a vaccine indicator ({\bf center}), and vaccine-year interaction ({\bf right}) including posterior means (cyan), 95\% pointwise credible intervals (light gray), 95\% simultaneous credible bands (dark gray), and a horizontal line at zero (dashed black). There is a substantial linear year-to-year decline in expected measles counts following the introduction of vaccine. This figure appears in color in the electronic version of this article, and color refers to that version. 
 \label{fig:coef}}
\end{center}
\end{figure}

\begin{figure}[h]
\begin{center}
\includegraphics[width=.49\textwidth]{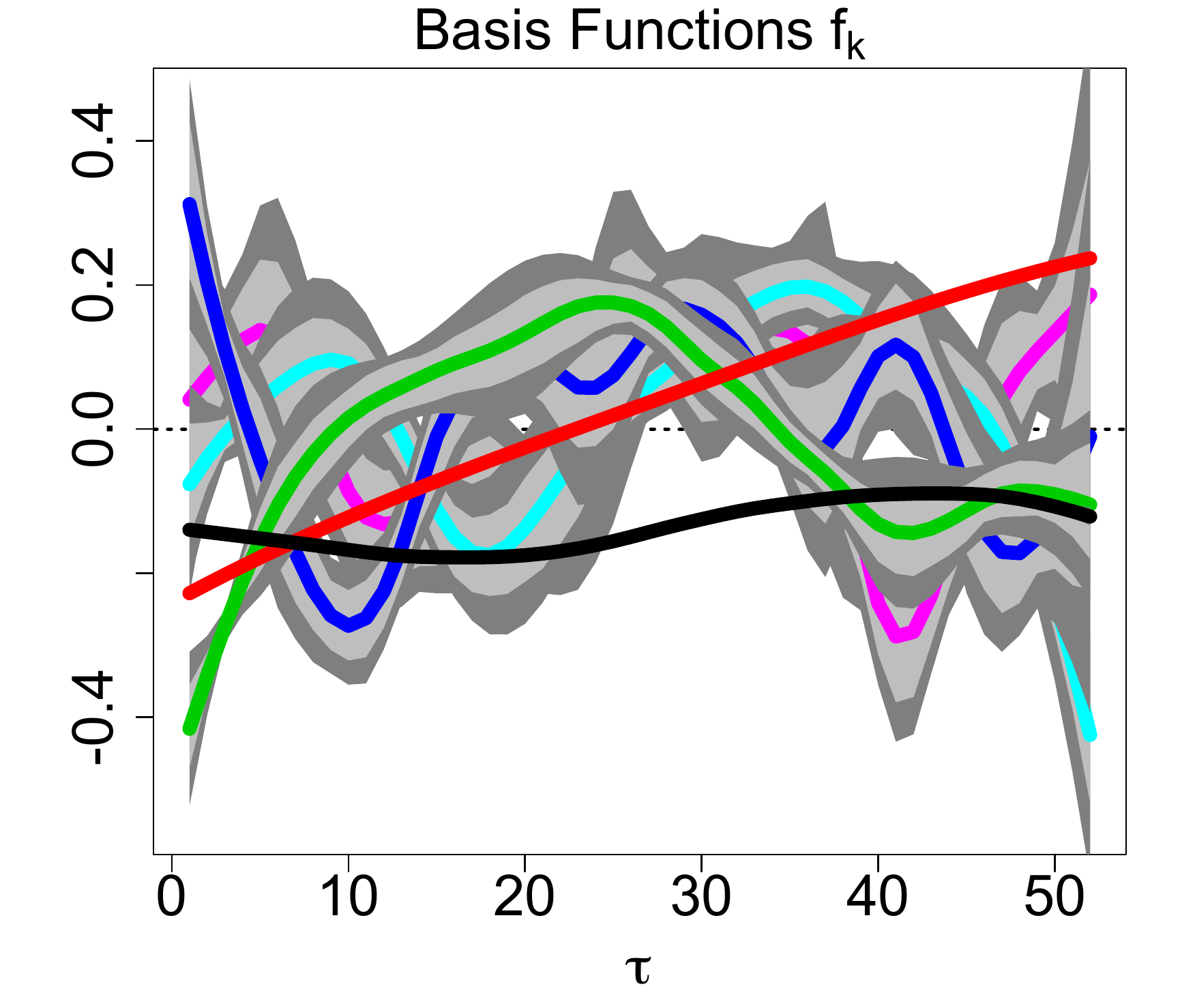}
\includegraphics[width=.49\textwidth]{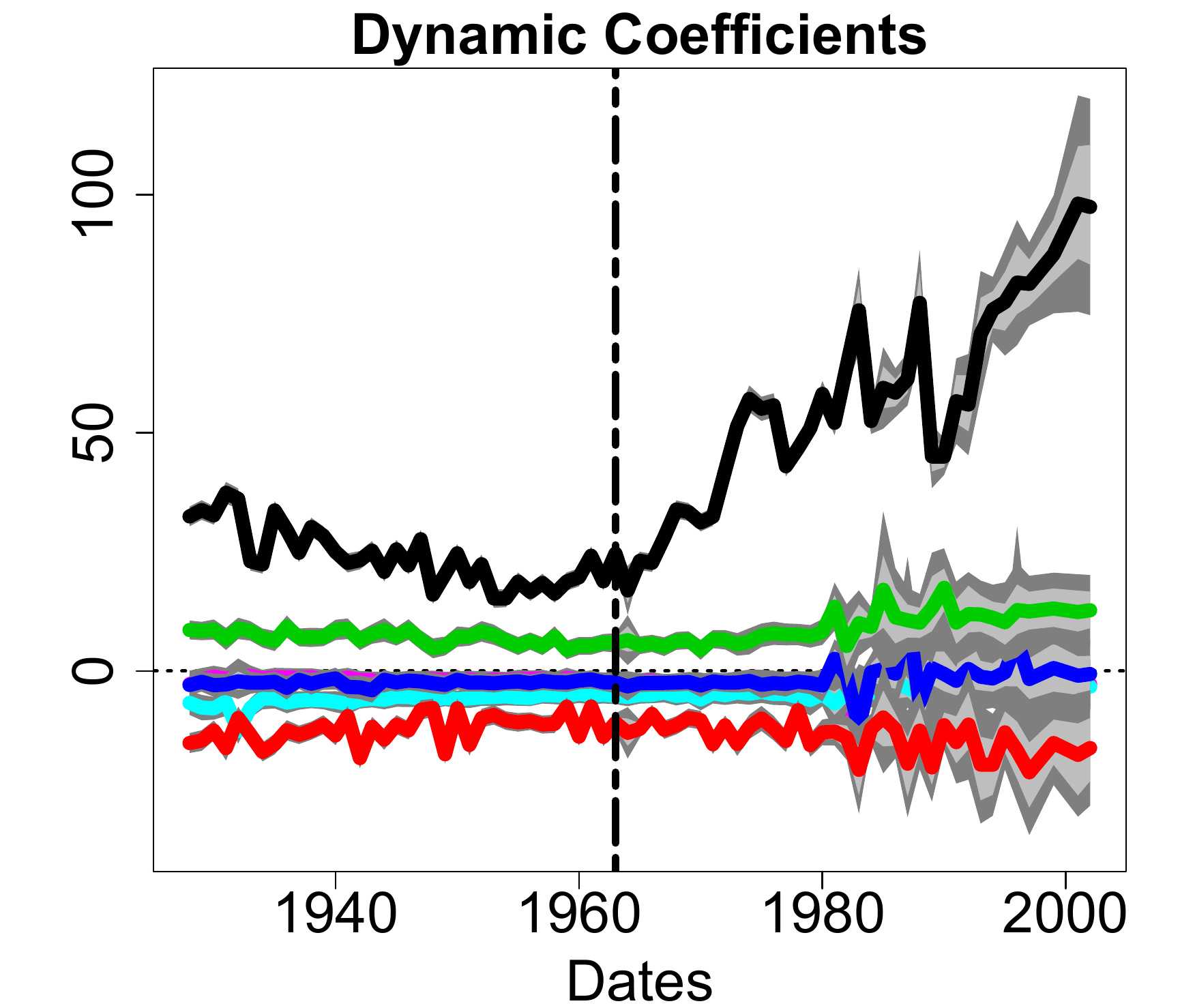}
\caption{Basis functions $\{f_k(\tau)\}$ ({\bf left}) and corresponding dynamic coefficients $\{\beta_{k,i}\}$ ({\bf right}) including posterior means (solid lines), 95\% pointwise credible intervals (light gray), and 95\% simultaneous credible bands (dark gray). The vertical dashed line (right) denotes the introduction of the vaccine in 1963. The increasing uncertainty among the coefficients $\{\beta_{k,i}\}$ after 1980 is likely due to an increase in missing data. This figure appears in color in the electronic version of this article, and color refers to that version. 
 \label{fig:flcs-factors}}
\end{center}
\end{figure}

Inference for the regression coefficients $\tilde \alpha_j(\tau)$ is in Figure \ref{fig:coef}, including posterior means and 95\% posterior pointwise intervals and simultaneous bands. There is a slightly positive pre-vaccine linear year-to-year trend, particularly in the fall. More importantly, there is a substantial linear year-to-year decline in measles counts post-vaccine, with the largest effect during the peak months of March-June. These effects are visually confirmed in Figure \ref{fig:flcs-factors}, which displays the learned seasonalities via $f_k$ and the year-to-year weights $\beta_{k,i}$. The dynamic coefficients $\beta_{k,i}$ show a slight linear trend pre-vaccine and a substantial trend in the oppose direction post-vaccine, especially for $k=1$. Interestingly, the learned basis functions $f_1$ and $f_2$ are simple: $f_1$ is sinusoidal and $f_2$ is linear. However, $f_k$ for $k>2$ are more complex, with $f_3$ partly capturing the peak months. Note that the coefficients $\{\beta_{k,i}\}$ for $k \le 4$ are visibly distinct from zero, suggesting that the more complex $f_k$ curves are important.

\section{Simulations}\label{sims}
We conducted a simulation study to validate the results from the forecasting comparison and evaluate the model performance under different distributions. We simulate data according to equations \eqref{nb-model}, \eqref{process}, \eqref{fdlm}, and \eqref{far} for $n = 50$ integer-valued functions observed at $m=50$  points, and introduce missingness for 10\% of the observations. For the  dispersion parameter in \eqref{nb-model}, we consider $r =1000$, which approximates the Poisson distribution, and $r=10$, which provides substantial overdispersion similar to the measles data.

For equally-spaced observation points $\tau \in [0,1]$, we introduce functional (or seasonal) dependence using the basis $f_1^*(\tau) = 1/\sqrt{m}$ and $f_k^*$ an orthogonal polynomial of degree $k$ for $k = 2,3,4$. For time-dependence, let $\beta_{k,i}^* = \sqrt{m} + \frac{\sqrt{m}}{k}\gamma_{k,i}^*$ with $\gamma_{k,i}^* = 0.8 \gamma_{k,i-1}^* +\eta_{k,i}^*$ and $\eta_{k,i}^* \stackrel{iid}{\sim}N(0, \sqrt{1 - 0.8^2})$ for $i=1,\ldots,n=50$. The real-valued process is simulated as $\theta_i^*(\tau) = \mu_i^*(\tau) + \sigma^*\epsilon_i^*(\tau)$, where $\mu_i^*(\tau) = \sum_{k=1}^4 f_k^*(\tau)\beta_{k,i}^*$, $\sigma^* = \mbox{sd}[\mu_i^*(\tau)]/\mbox{RSNR}$ for root signal-to-noise ratio RSNR = 10, and $\epsilon_i^*(\tau) \stackrel{iid}{\sim}N(0,1)$. The observed integer-valued functional data are simulated from $Z_i(\tau) \stackrel{indep}{\sim} \mbox{NB}\left(r,  \exp\left[\theta_i^*(\tau)\right]/\left\{r + \exp\left[\theta_i^*(\tau)\right]\right\}\right)$, and 10\% of the observations $Z_i(\tau)$ are deleted at random to induce missingness. We define the true curves to be the conditional expectations $Z_i^*(\tau) = \exp\left[\mu_i^*(\tau)\right]\exp\left[(\sigma^*)^2/2\right]$ akin to \eqref{cond-mean-2}. Given all observed counts $\{Z_i(\tau\}_{i=1}^{n-1}$ prior to time $n$ and the first 30 observation points at time $n$ $\{Z_n(\tau_j)\}_{j=1}^{30}$, the goal is to forecast the true curves $\{Z_n^*(\tau_j)\}_{j=31}^{50}$ for the remaining 20 observation points. The same competing methods are considered as in Section \ref{measles:fore} and the simulations are repeated 100 times. Note that empirical coverage probability (ECP) is evaluated based on future counts $\{Z_n(\tau_j)\}_{j=31}^{50}$ rather than the true curves $\{Z_n^*(\tau_j)\}_{j=31}^{50}$. 

\begin{figure}[h]
\begin{center}
\includegraphics[width=.49\textwidth]{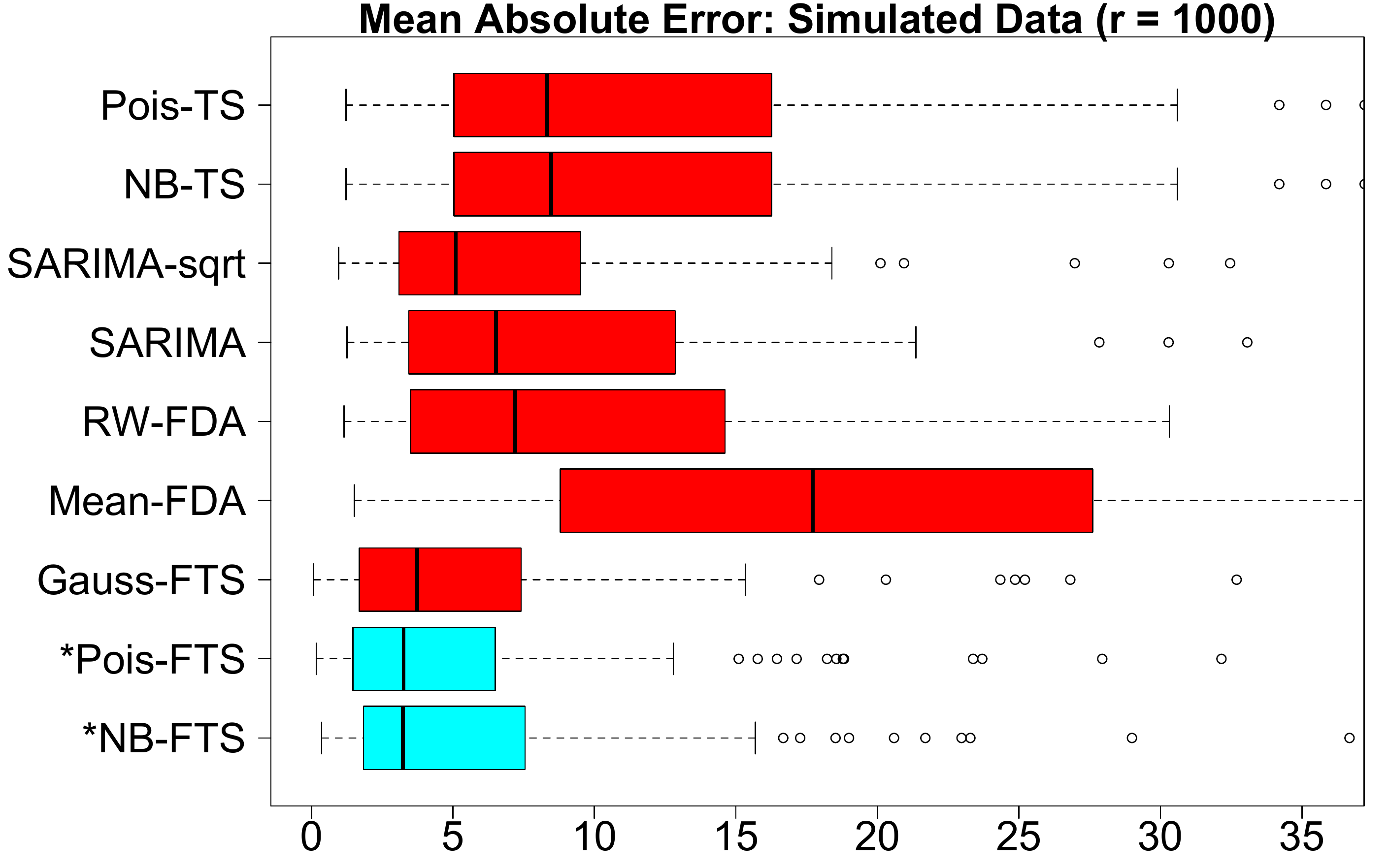}
\includegraphics[width=.49\textwidth]{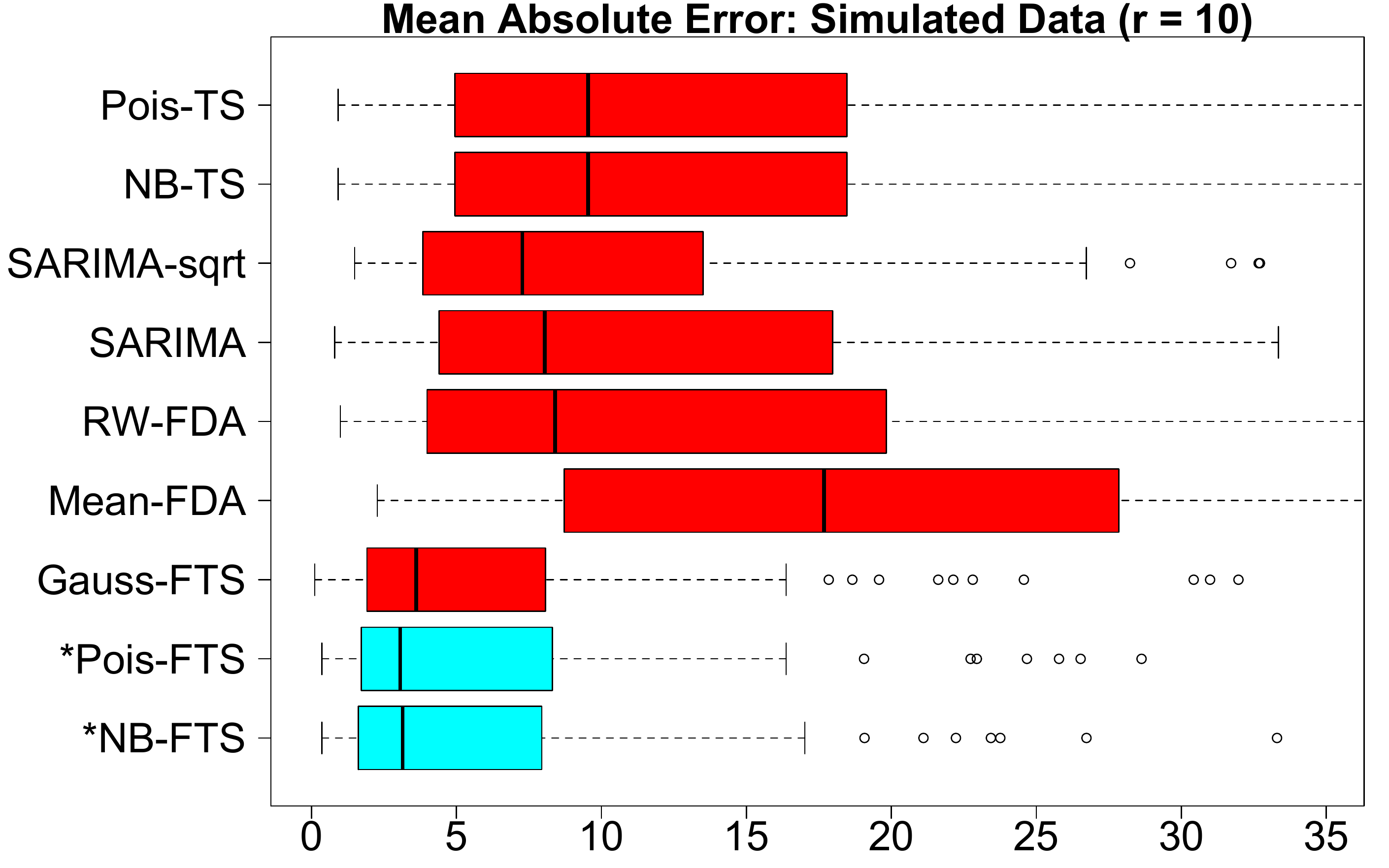}

\includegraphics[width=.49\textwidth]{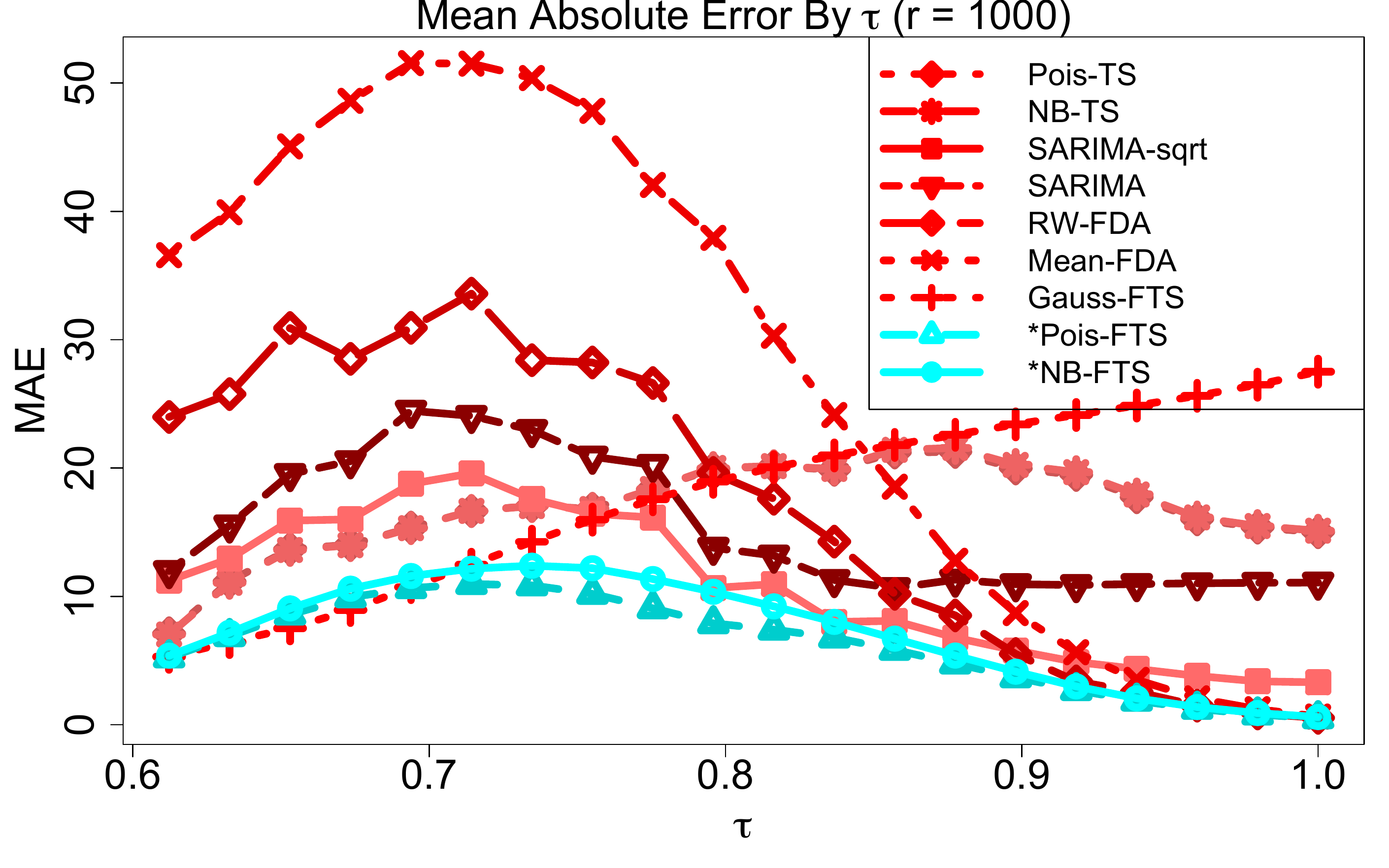}
\includegraphics[width=.49\textwidth]{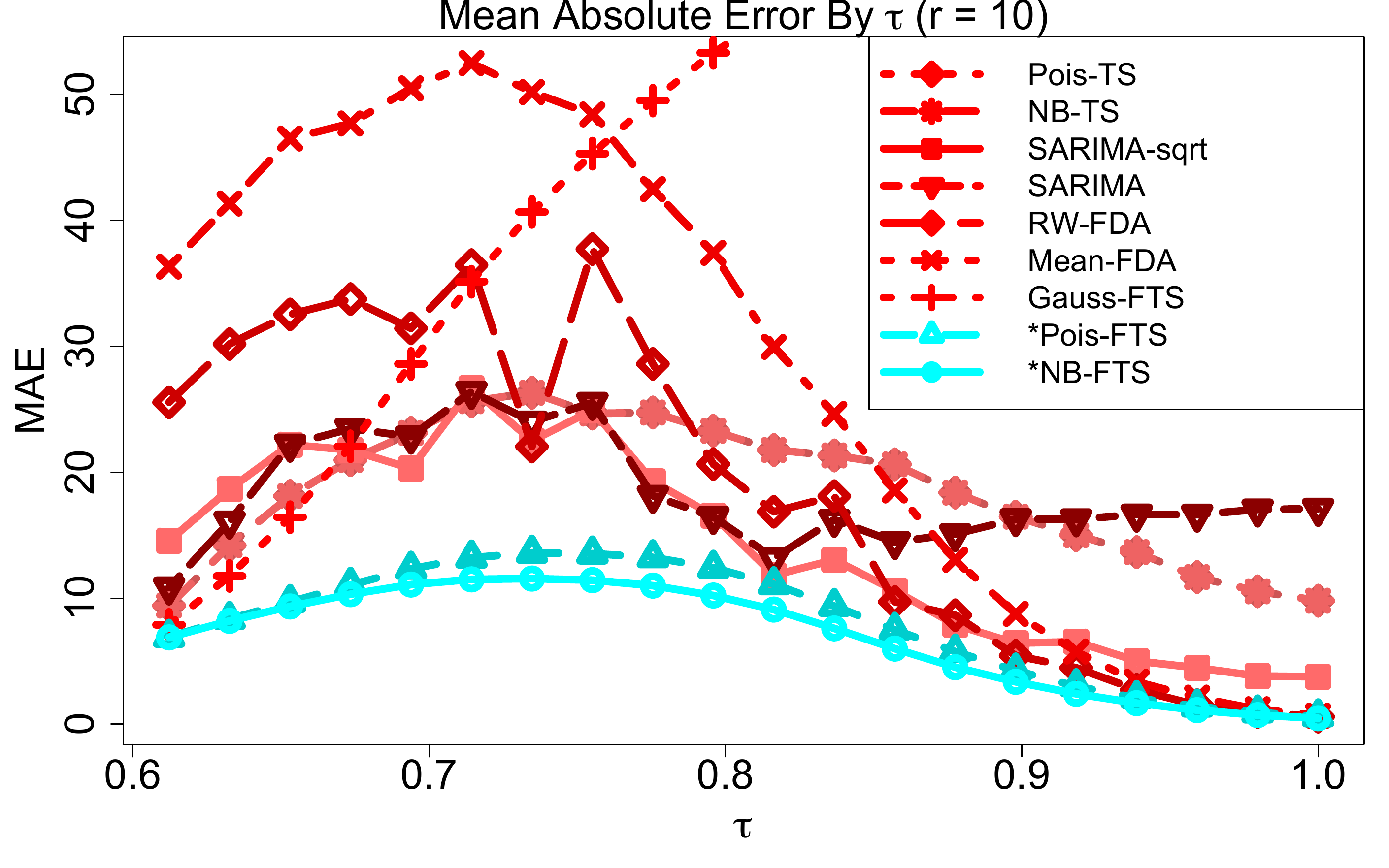}
\caption{Mean absolute error for $r=1000$ ({\bf left}) and $r=10$ ({\bf right}) for forecasting the remaining time points 31-50. {\bf Top:} $\mbox{MAE}_i$ across simulations $i$ and aggregated over $\tau^*$. {\bf Bottom:} $\mbox{MAE}(\tau^*)$ for each $\tau^*$ and aggregated by simulation. The proposed methods *NB-FTS and *Pois-FTS (cyan) perform best in both settings. Note that extreme outliers from competing methods were omitted for display purposes. This figure appears in color in the electronic version of this article, and color refers to that version. 
 \label{fig:mae-sim}}
\end{center}
\end{figure}

\begin{figure}[h]
\begin{center}
\includegraphics[width=.8\textwidth]{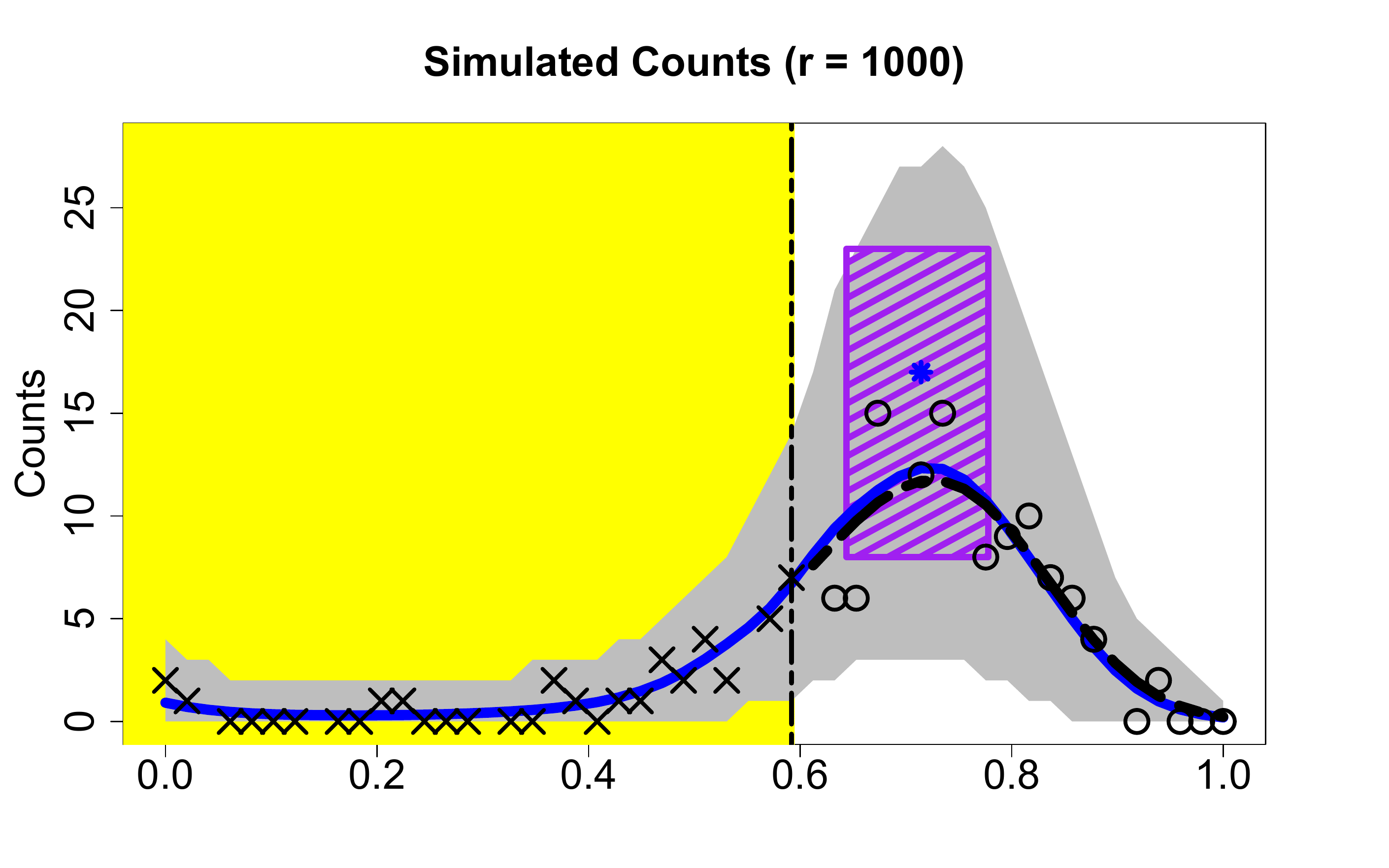}
\includegraphics[width=.4\textwidth]{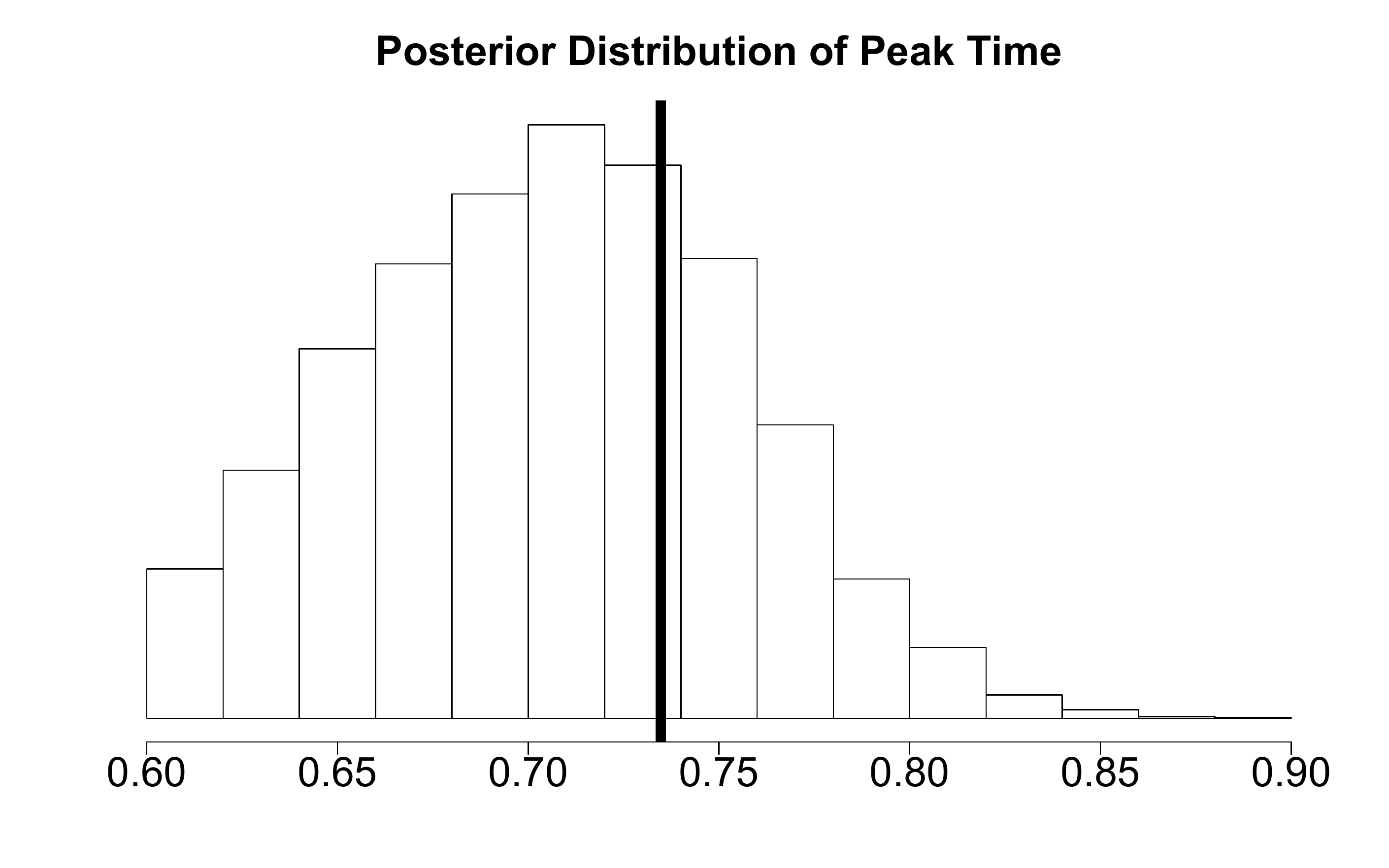}
\includegraphics[width=.4\textwidth]{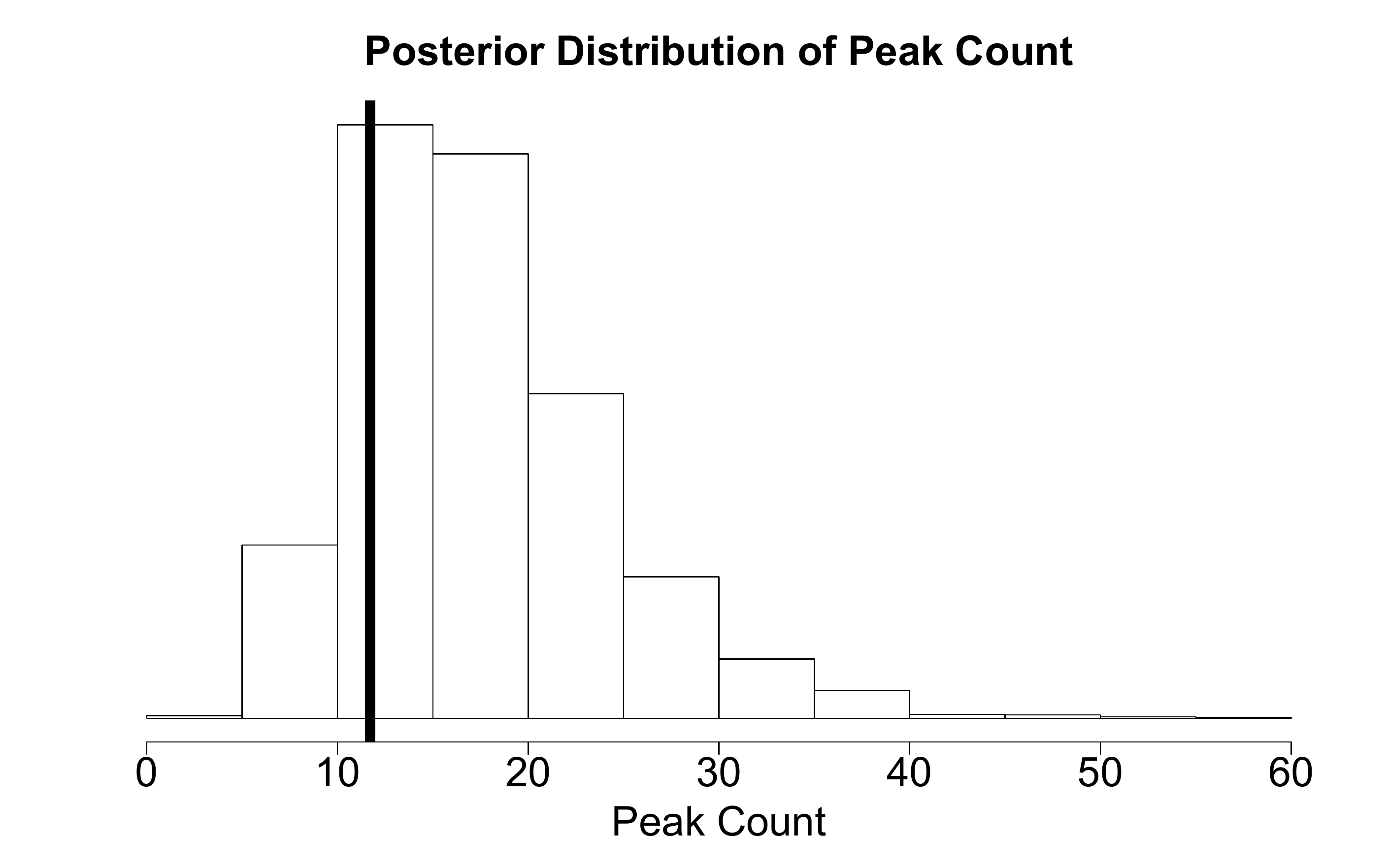}
\caption{An identical format as Figure \ref{fig:emp10-52}, but for simulated data with $r=1000$. {\bf Top:} Using data from only the first 30 time points (x-marks, yellow region), we forecast the counts for the remaining 20 time points (circles). The posterior predictive mean (blue line) and 95\% intervals (gray region) provide a forecast with uncertainty quantification, which closely follows the true curve (dashed black line). The posterior predictive distribution provides inference for the peak time and peak counts: here, the posterior median time and peak count (blue star) and 80\% posterior credible intervals for the peak time and peak count (purple region), as well as the posterior distribution for the peak time ({\bf bottom left}) and the posterior distribution for the peak count ({\bf bottom right}) together with the true values (vertical black lines, {\bf bottom}). 
\label{fig:emp-sim1000}}
\end{center}
\end{figure}

\begin{figure}[h]
\begin{center}
\includegraphics[width=.8\textwidth]{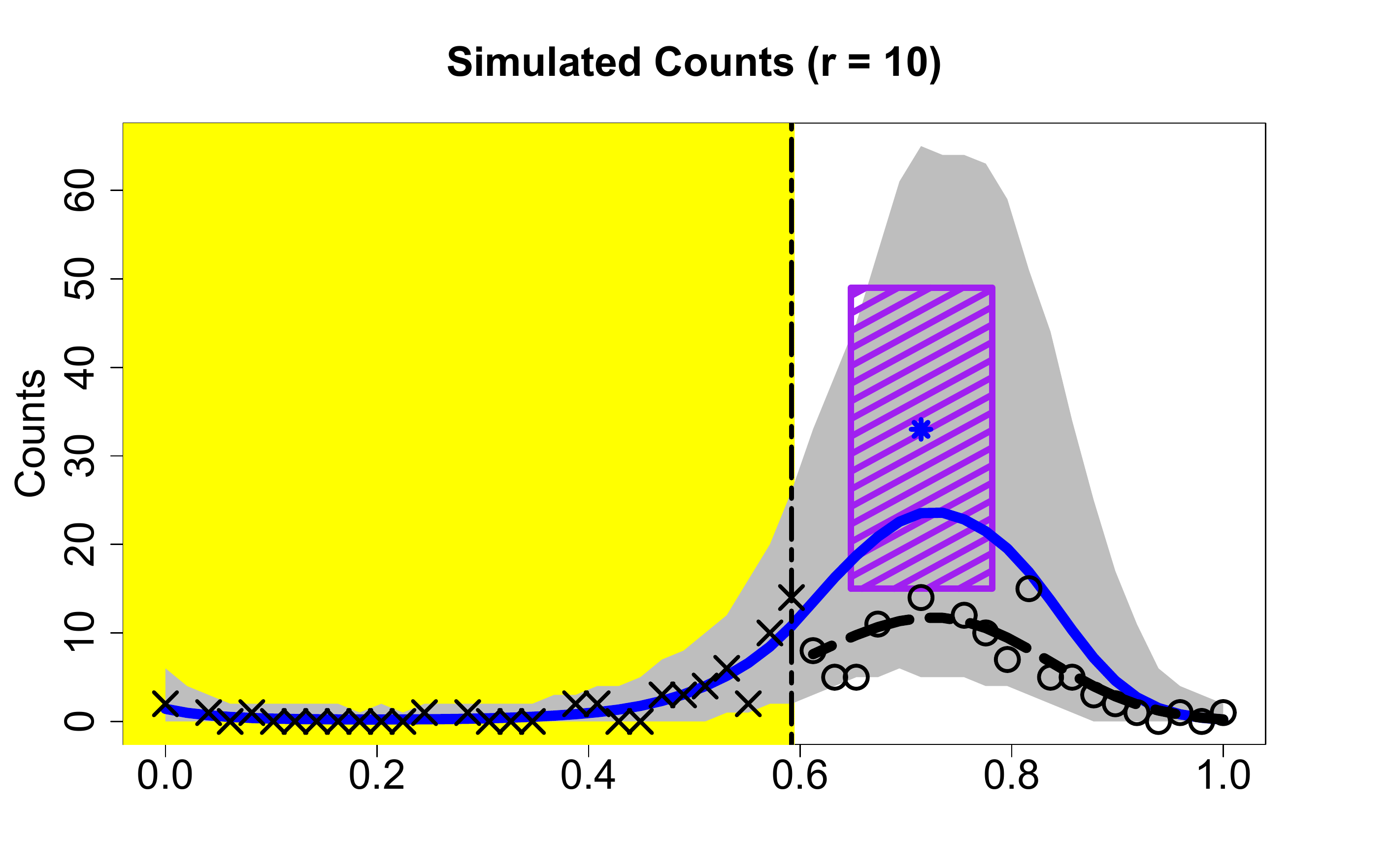}
\includegraphics[width=.4\textwidth]{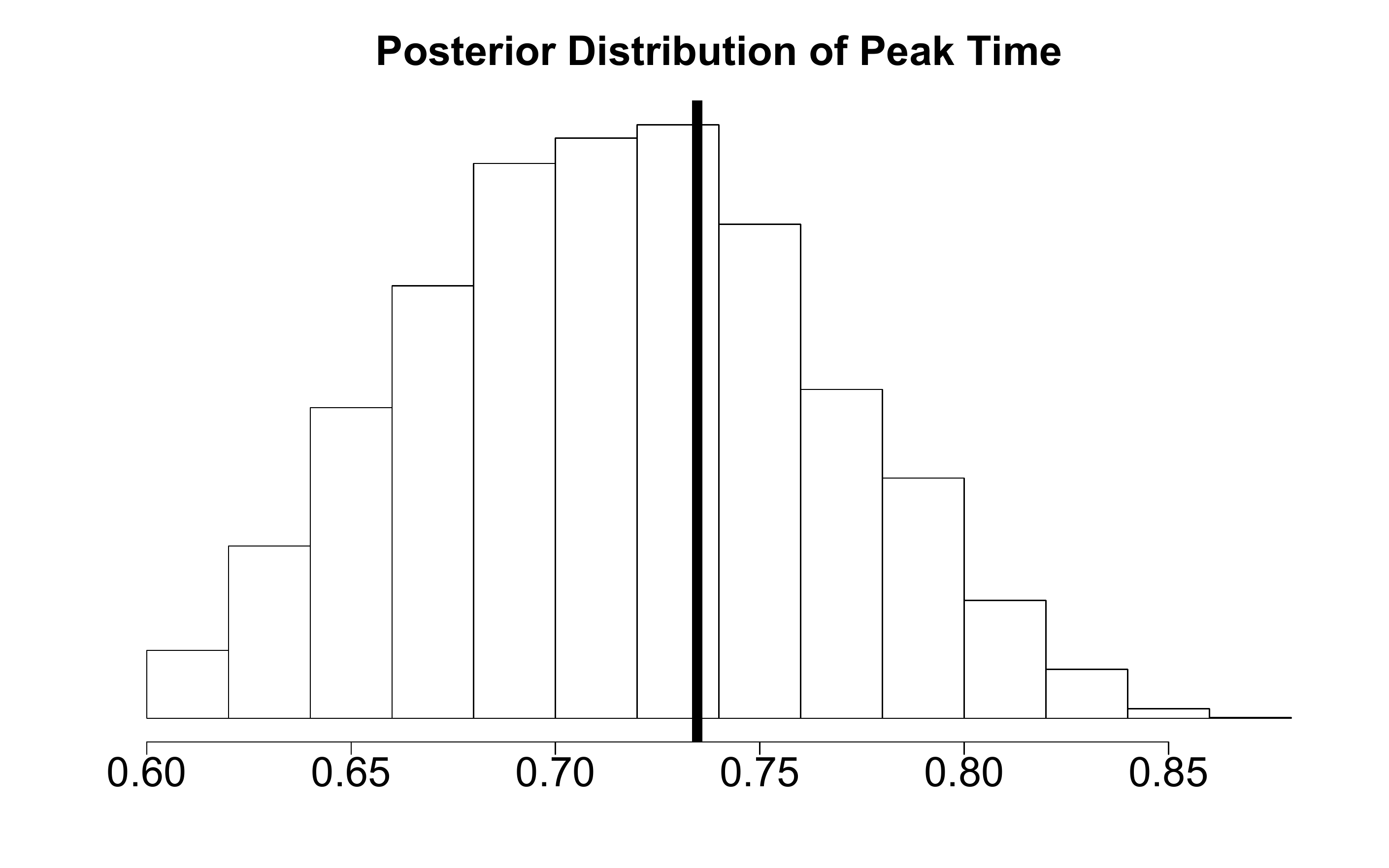}
\includegraphics[width=.4\textwidth]{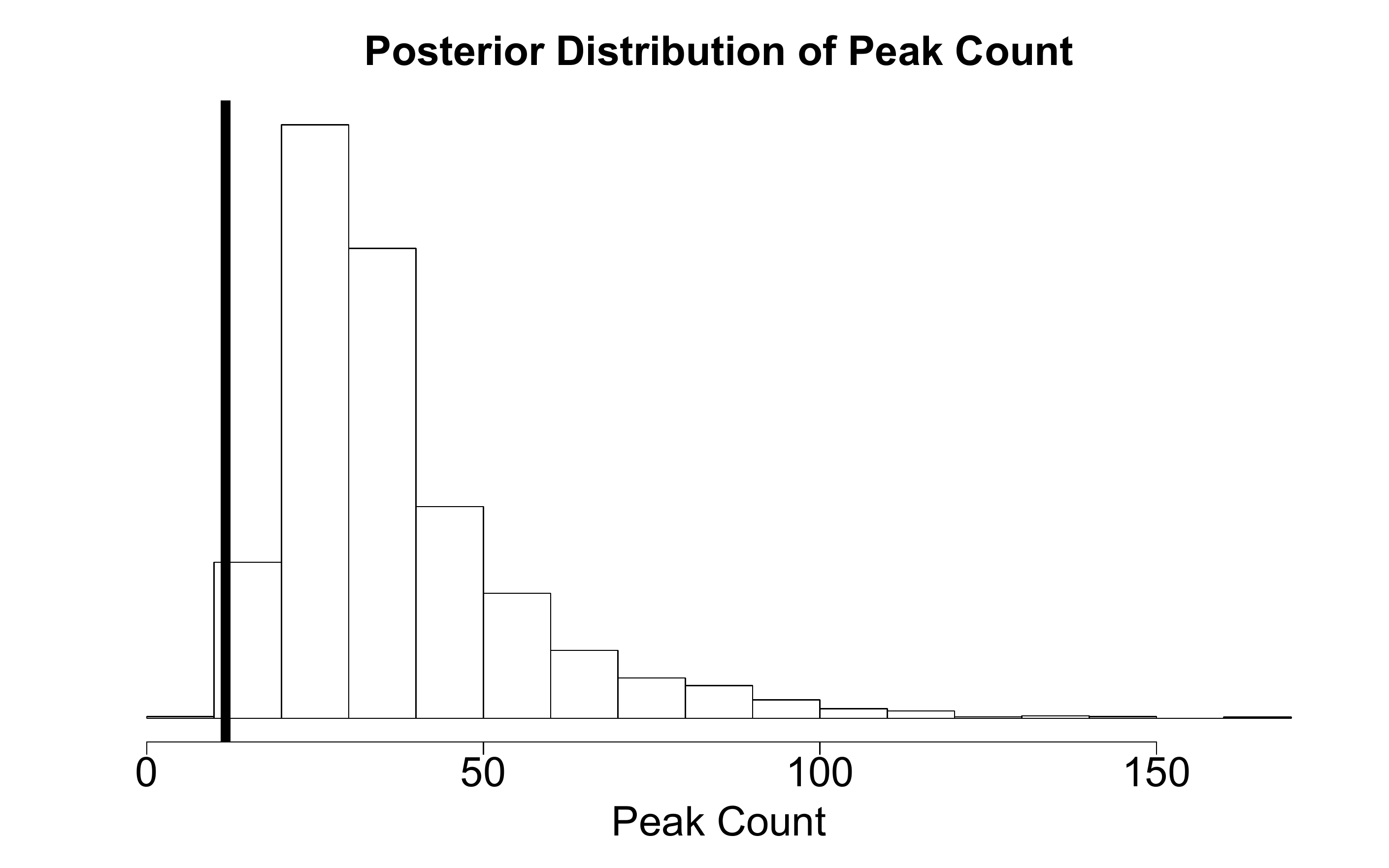}
\caption{An identical format as Web Figure \ref{fig:emp-sim1000}, but with $r=10$. With larger variance and overdispersion, forecasting is more challenging: the mean posterior predictive distribution (blue line) overestimates the true curve (dashed black line), as does the posterior predictive distribution of the peak count ({\bf bottom right}). 
 \label{fig:emp-sim10}}
\end{center}
\end{figure}

An illustration of the forecasting design, analogous to Figure \ref{fig:emp10-52}, is provided in Web Figure \ref{fig:emp-sim1000} for $r=1000$ and Web Figure \ref{fig:emp-sim10} for $r=10$. The point and interval forecasts perform well in both cases, although the $r=10$ case is far more challenging due to the increased variability. Aggregating across all simulations, the MAEs defined in Section \ref{measles:fore} are displayed in Figure \ref{fig:mae-sim}. The functional models NB-FTS, Pois-FTS, and Gauss-FTS perform similarly for $\mbox{MAE}_i$, while the proposed integer-valued methods NB-FTS and Pois-FTS  offer clear improvements in $\mbox{MAE}(\tau^*)$ across all forecast horizons. Figure \ref{fig:cover-sim} provides ECPs and MIWs for comparing forecast intervals. As for the measles data, NB-FTS achieves the correct nominal coverage with narrower intervals than competing methods. Notably, Gauss-FTS produces similar interval widths, yet suffers from substantial undercoverage. Lastly, coverage for the peak count and peak time are in Table \ref{table:cover-epi}, with NB-FTS and Pois-FTS again outperforming Gauss-FTS. Clearly, the integer-valued distribution is important for well-calibrated and precise uncertainty quantification.


\section{Discussion}\label{discussion}
The proposed framework for integer-valued functional data offers clear improvements in modeling and forecasting measles counts relative to existing methods. The methodology is fully Bayesian with an integer-valued distribution for the observed data, which produces correctly calibrated and accurate uncertainty quantification for measles counts and epi-relevant features, including the peak number of counts and the time at which the peak occurs. The intra-year seasonality is learned from the data, and includes  year-to-year dynamic dependence. Importantly, the model is sufficiently flexible and accurate in both pre- and post-vaccine eras, and therefore offers utility in the modern era in which vaccines are available yet vaccination rates are declining rapidly. More broadly, the model and accompanying Gibbs sampler are designed to build upon methods and algorithms for Gaussian functional data, such as functional autoregressions and function-on-scalar regressions, which suggests a widely generalizable modeling framework.

Given the success of the proposed methodology, there are a number of promising extensions. First, we may consider disease counts for multiple related diseases or multiple states concurrently, which requires a \emph{multivariate} functional data model \citep{kowal2017bayesian}. Similarly, external data sources such as Google search trends, Twitter, or Wikipedia have proven useful for influenza forecasting \citep{paul2014twitter,hickmann2015forecasting}, and may be incorporated for measles forecasting via the regression model \eqref{far-reg} or suitable modifications. Lastly, methodological extensions for binomial functional data are readily available via a similar P{\'o}lya-Gamma data augmentation scheme. 



\bibliographystyle{apalike}
\bibliography{refs}

\section*{Supporting Information}
Additional supporting information may be found online in the Supporting Information section at the end of the article, including Web Figures, Tables, and \texttt{R} code. 

\begin{table}[ht]
\centering
\begin{tabular}{rrrr}
  \hline
Measles Peak Value (10-52)  & *NB-FTS & *Pois-FTS & Gauss-FTS \\ 
  \hline
Median Width & {\bf 2535} & 1504 & 2717 \\ 
 Coverage Probability & {\bf 0.97} & 0.84 & 0.68 \\ 
   \hline
   
    \hline
    
    \hline
Measles Peak Time (10-52) & *NB-FTS & *Pois-FTS & Gauss-FTS \\ 
  \hline
Median Width & 11 & {\bf 10} & 39 \\ 
 Coverage Probability & 0.94 & {\bf 1.00} & 0.90 \\ 
   \hline
   
    \hline
    
    \hline
   Simulated Peak Value ($r=1000$)  & *NB-FTS & *Pois-FTS & Gauss-FTS \\ 
  \hline
Median Width & {\bf 50} & 41 & 42 \\ 
 Coverage Probability & {\bf 0.95} & 0.94 & 0.88 \\ 
     \hline
   
    \hline
    
    \hline
    Simulated Peak Time ($r=1000$)   & *NB-FTS & *Pois-FTS & Gauss-FTS \\ 
  \hline
Median Width & 16 & {\bf 14} & 16 \\ 
 Coverage Probability & 1.00 & {\bf 1.00} & 1.00 \\ 
        \hline
   
    \hline
    
    \hline 
    Simulated Peak Value ($r=10$)     & *NB-FTS & *Pois-FTS & Gauss-FTS \\ 
  \hline
Median Width & 74 & {\bf 48} & 49 \\ 
 Coverage Probability & 0.96 & {\bf 0.96} & 0.86 \\ 
          \hline
   
    \hline
    
    \hline 
    Simulated Peak Time ($r=10$)      & *NB-FTS & *Pois-FTS & Gauss-FTS \\ 
  \hline
Median Width & 16 & {\bf 14} & 18 \\ 
 Coverage Probability & 1.00 & {\bf 1.00} & 0.99 \\ 
   \hline
\end{tabular}
\caption{Median interval width and percent coverage for 95\% forecasting intervals of peak time and peak value. The results are given for measles forecasting of  weeks 10-52 and the simulated data with $r=1000$ and $r=10$. The narrowest intervals achieving 95\% coverage are in bold. The proposed integer-valued functional data models *NB-FTS and *Pois-FTS clearly outperform the Gaussian functional data model. *NB-FTS typically achieves the correct coverage, although in some cases produces wider intervals. Other methods are omitted, since they do not readily admit posterior predictive intervals for epi-relevant features.}
\label{table:cover-epi}
\end{table}

\clearpage 

\captionsetup[figure]{name={Web Figure}}
\captionsetup[table]{name={Web Table}}

\section*{Web Appendix A}

\begin{table}[ht]
\centering
\begin{tabular}{rrrrrrrr}
        \hline
Pre-vaccine (10-52) & *NB-FTS & *Pois-FTS & Gauss-FTS & SARIMA & SARIMA-sqrt & NB-TS & Pois-TS \\ 
  \hline
Median Width & {\bf 543} & 444 & 1117 & 2947 & 1751 & 600 & 210 \\ 
 Coverage Probability & {\bf 0.96} & 0.84 & 0.89 & 0.79 & 0.92 & 0.80 & 0.13 \\ 
    \hline
 Post-vaccine (10-52) & *NB-FTS & *Pois-FTS & Gauss-FTS & SARIMA & SARIMA-sqrt & NB-TS & Pois-TS \\ 
  \hline
Median Width & {\bf 78} & 72 & 319 & 3759 & 1019 & 37 & 142 \\ 
 Coverage Probability & {\bf 0.98} & 0.94 & 0.89 & 0.99 & 1.00 & 0.69 & 0.08 \\ 
   \hline
   
    \hline
    
    \hline
    Pre-vaccine (26-52) & *NB-FTS & *Pois-FTS & Gauss-FTS & SARIMA & SARIMA-sqrt & NB-TS & Pois-TS \\ 
  \hline
Median Width & {\bf 244} & 202 & 353 & 2938 & 1189 & 729 & 154 \\ 
 Coverage Probability & {\bf 0.96} & 0.87 & 0.97 & 1.00 & 0.99 & 0.94 & 0.09 \\ 
   \hline
Post-vaccine (26-52) & *NB-FTS & *Pois-FTS & Gauss-FTS & SARIMA & SARIMA-sqrt & NB-TS & Pois-TS \\ 
  \hline
Median Width & {\bf 32} & 33 & 100 & 3707 & 867 & 36 & 70 \\ 
 Coverage Probability&{\bf  0.95} & 0.92 & 0.95 & 1.00 & 1.00 & 0.82 & 0.30 \\ 
   \hline
   
    \hline
    
    \hline
Simulated ($r = 1000$) & *NB-FTS & *Pois-FTS & Gauss-FTS & SARIMA & SARIMA-sqrt & NB-TS & Pois-TS \\ 
  \hline
Median Width & {\bf 20} & 19 & 22 & 73 & 33 & 49 & 24 \\ 
 Coverage Probability & {\bf 0.96} & 0.93 & 0.77 & 0.92 & 0.92 & 0.93 & 0.61 \\ 
  \hline
Simulated ($r = 10$)  & *NB-FTS & *Pois-FTS & Gauss-FTS & SARIMA & SARIMA-sqrt & NB-TS & Pois-TS \\ 
  \hline
Median Width & {\bf 24} & 20 & 26 & 97 & 42 & 42 & 21 \\ 
 Coverage Probability & {\bf 0.96} & 0.91 & 0.77 & 0.92 & 0.90 & 0.94 & 0.56 \\ 
   \hline
\end{tabular}
\caption{Median interval width and percent coverage for 95\% forecasting intervals. The results are separated by weeks forecasted (10-52 or 26-52) and pre-vaccine (before 1963) and post-vaccine (after 1964) for the measles data, and by $r=1000$ and $r=10$ for the simulated data. The narrowest intervals achieving 95\% coverage are in bold. Intervals are substantially wider pre-vaccine. Notably, only the proposed method *NB-FTS is capable of simultaneously achieving the nominal 95\% coverage while maintaining narrow intervals.}
\label{table:cover}
\end{table}

\end{document}